\documentclass[11pt]{article}
\usepackage{latexsym,amsmath,amsthm,amsfonts,amsbsy,amscd}
\newtheorem{lemma}{Lemma}

\def\Integer{\mathbb{Z}}
\newcommand{\Mult}[2]{\genfrac{[}{]}{0pt}{}{#1}{#2}}
\def\mod#1{\; (\bmod \: #1)}
\def\I{{\cal I}}
\def\bs{\boldsymbol}

\def\vn{\bs n}
\def\vm{\bs m}
\def\vu{\bs u}
\def\vv{\bs v}
\def\vA{\bs A}
\def\vQ{\bs Q}
\def\vet{\bs \eta}
\def\vmu{\bs \mu}
\def\vv{\bs v}
\def\ve{\bs e}

\def\vi{\bs i}
\def\vE{\bs E}

\def\vbeta{\bs \beta}

\def\ce#1{\theta(#1~{\rm even})}
\def\co#1{\theta(#1~{\rm odd})}
\def\dto{\stackrel{d}{\to}}

\def\th{\hat{t}}
\def\uh{\hat{u}}
\def\Ah{\hat{A}}
\def\Qh{\hat{Q}}
\def\lh{\hat{l}}
\def\nuh{\hat{\nu}}
\def\nh{\hat{n}}
\def\mh{\hat{m}}
\def\Bh{\hat{B}}
\def\vmh{\hat{\vm}}
\def\vnh{\hat{\vn}}
\def\vAh{\hat{\vA}}
\def\vuh{\hat{\vu}}
\def\vQh{\hat{\vQ}}

\def\fh{\hat{f}}
\def\gh{\hat{g}}
\def\yh{\hat{y}}
\def\yhb{\hat{\bar{y}}}
\def\lh{\hat{l}}

\def\tf{\tilde{f}}

\def\bl{\bar{l}}
\def\by{\bar{y}}

\def\srs#1#2#3{\sum_{\substack{#1\in\Integer_+^{N-1}\\
\frac{#2}{N}-(C^{-1}#1)_1 \in \Integer+\frac{#3}{2}}}}

\numberwithin{equation}{section}
\numberwithin{lemma}{section}
\numberwithin{corollary}{section}
\numberwithin{definition}{section}
\numberwithin{theorem}{section}

\begin{document}

\title{Bailey flows and Bose--Fermi identities for the conformal \\ 
coset models  $({\rm A}^{(1)}_1)_N\times ({\rm A}^{(1)}_1)_{N'}/ 
({\rm A}^{(1)}_1)_{N+N'}$}

\date{February 1997}

\author{
\Large Alexander Berkovich\thanks{
e-mail: {\tt berkov\_a@math.psu.edu}}\\
\mbox{}\\
\em Department of Mathematics, The Pennsylvania State University \\
\em University Park, PA 16802, USA \\
\mbox{}\\
\Large Barry M. McCoy\thanks{
e-mail: {\tt mccoy@insti.physics.sunysb.edu}} \ 
\Large and Anne Schilling\thanks{
e-mail: {\tt anne@insti.physics.sunysb.edu}} \\
\mbox{} \\
\em Institute for Theoretical Physics, State University of New York \\
\em Stony Brook, NY 11794-3840, USA \\
\mbox{} \\
\Large S.~Ole Warnaar\thanks{
e-mail: {\tt warnaar@maths.mu.oz.au}} \\
\mbox{} \\
\em Department of Mathematics, University of Melbourne\\
\em Parkville, Victoria 3052, Australia}
\maketitle

\begin{abstract}
We use the recently established higher-level Bailey lemma and Bose--Fermi 
polynomial identities for the minimal models $M(p,p')$ to demonstrate the 
existence of a Bailey flow from $M(p,p')$ to the coset models 
$({\rm A}^{(1)}_1)_N\times ({\rm A}^{(1)}_1)_{N'}/({\rm A}^{(1)}_1)_{N+N'}$ 
where $N$ is a positive integer and $N'$ is fractional, and to obtain
Bose--Fermi identities for these models. The fermionic side of these
identities is expressed in terms of the fractional-level Cartan matrix
introduced in the study of $M(p,p')$. Relations 
between Bailey and renormalization group flow are discussed. 
\end{abstract}

\section{Introduction} 

A decade ago A. Zamolodchikov~\cite{ZAMA} and Ludwig and
Cardy~\cite{LC} studied the phenomena of
renormalization group (RG) flow from the minimal model $M(p,p+1)$ to the
model $M(p-1,p)$ by means of perturbation theory when $p$ is very large.
These flows define a one parameter family of massless field
theories. In 1991 Al. Zamolodchikov~\cite{ALZAM} studied the ground
state energy (or, equivalently, the effective central charge) for the one
parameter flow from  $M(4,5)$ to $M(3,4)$ in terms of the
thermodynamic Bethe Ansatz equations and conjectured the
generalization to all $M(p,p+1).$
Since then many other flows have been discovered using this pertubative 
method, such as the flow from $M(p,p')$ to $M(2p-p',p)$~\cite{AHN} and 
flows between coset models~\cite{CSS,Zam91}.

All of these studies have been made using the techniques and philosophy
of the renormalization group and all of them illustrate the famous
$c$ theorem~\cite{ZAMB} by flowing from a larger to a smaller
effective central charge. For this reason it is often stated that 
renormalization group flow is irreversible.

It is thus most interesting that recently a nonperturbative construction from 
the mathematical literature of the Rogers--Ramanujan identities was 
used~\cite{FQ94,Ole2,BMS96a} to make connection 
between the minimal models in the opposite direction from the
renormalization group flow. 
This construction is referred to as {\em Bailey flow}.

Bailey flow originates in the work of Bailey~\cite{Bailey}
and Slater \cite{Slater51,Slater52} in their proofs of the many $q$-series 
identities of Rogers \cite{Rogers94,Rogers17} and has been greatly extended by 
Andrews \cite{Andrews75,Andrews84} and others~\cite{Paule}-\cite{SW96hl2}.
The prototypical $q$-series identity is the original identity of
Rogers and Ramanujan~\cite{Rogers94,Ram},
\begin{equation}\label{rr}
\frac{1}{(q)_{\infty}}\sum_{j=-\infty}^{\infty}
\Bigl(q^{j(10j+1+2a)}-q^{(2j+1)(5j+2-a)}\Bigr)=
\sum_{n=0}^{\infty}\frac{q^{n(n+a)}}{(q)_n},
\end{equation}
where
$a=0,1$, and
\begin{equation} \label{q_symb}
(x;q)_n = (x)_n = (1-x)(1-xq) \cdots (1-xq^{n-1})
\quad \text{for } n\geq 1 \quad \text{and} \quad (x)_0=1.
\end{equation}
The left-hand side is recognized as the character of the $M(2,5)$
minimal model in the form of~\cite{Dobrev,CIZ87} which is computed using the 
construction of Feigin and Fuchs~\cite{FF83}. 
This construction involves the elimination of states from a
bosonic Fock space. Thus we refer to the left-hand side as a bosonic
representation. Conversely, the right-hand side has an
interpretation in terms of fermionic quasi-particles~\cite{KKMM1,KKMM2} and is
called a fermionic representation. 

Bailey's lemma is a constructive proceedure which starts from  a polynomial
generalization of a Bose--Fermi identity for the characters/ branching
functions of the initial conformal field theory (CFT) and produces a 
Bose--Fermi equality for a rational generalization of the 
characters/branching functions of some other CFT
\begin{equation*}
\begin{CD}
\text{BOSE}=\text{FERMI}\\
@VVV{\text{Bailey's lemma}}  \\
\text{BOSE}'=\text{FERMI}'.
\end{CD}
\end{equation*}

The Bailey flow from $M(p-1,p)$ to $M(p,p+1)$ is discussed 
in~\cite{FQ94,Ole2,BMS96a} 
and further flows to the unitary $N=1$ supersymmetric model $SM(p,p+2)$
and to the $N=2$ supersymmetric models with $c=3(1-2/p)$ are given 
in~\cite{BMS96a}. 
The Bailey flow from $M(p-1,p)$ to $M(p,p+1)$ is in
the opposite direction from the RG flow and the remainder of the
Bailey flows give relations between CFT's which have not previously
been seen in the RG analysis. In a sense the Bailey construction is an
``Aufbau Prinzip'' and as such it promises to provide an alternative
construction for most (possibly all) conformal field theories.

In this paper we extend these ideas to find a flow from
the general minimal model $M(p,p')$ to the coset models
\begin{equation}
\frac{({\rm A}^{(1)}_1)_N \times ({\rm A}^{(1)}_1)_{N'}}
{({\rm A}^{(1)}_1)_{N+N'}}
\label{coset}
\end{equation}
for integer level $N$ and fractional level $N'$.
For brevity these coset theories will be denoted by $(P,P')_N$, where 
$$N'=\frac{N P}{P'-P}-2  \quad \text{or}  \quad N'=-2-\frac{N P'}{P'-P},$$
with the restrictions $P<P'$, $P'-P \equiv 0 \mod N$ and 
$\gcd(\tfrac{P'-P}{N},P')=1$.
In this notation the minimal models $M(p,p')$ correspond to the coset model
$(p,p')_1$ as first observed in~\cite{Kent}. 

Using the method of Feigin and Fuchs~\cite{FF83} the bosonic form of the 
characters was given explicitly in~\cite{RC83} for the unitary 
minimal models $M(p,p+1)$ and in~\cite{Dobrev,CIZ87} for the non-unitary 
cases $M(p,p')$.
The branching functions for the cosets~\eqref{coset}
for integer levels was given in~\cite{D87,D88} by computing configuration
sums of RSOS models and
in~\cite{KMQ}-\cite{R} using the Feigin and Fuchs construction.
Kac and Wakimoto~\cite{KW88} introduced admissible representations of affine 
Lie algebras which in general correspond to fractional levels and
non-unitary CFTs. This paved the way for the study of the coset 
models~\eqref{coset} with one fractional level~\cite{Ahn91} and
two fractional levels~\cite{ACT}.
Further aspects of coset~\eqref{coset}, including its spin content, have 
been studied in~\cite{BEHHH} in the context of unifying ${\cal W}$-algebras.

Our method to obtain the Bailey flow from $M(p,p')$ to the 
cosets~\eqref{coset} is to use the polynomial identities for $M(p,p')$ 
recently established in~\cite{BM96,BMS96} with the new extension of the
Bailey construction obtained in~\cite{SW96hl1,SW96hl2}. The branching function
identities obtained from this flow have previously been found for
the unitary case $(p,p+N)_N$~\cite{S96a,S96b}, for $(3,2N+3)_N$~\cite{W} 
and the cosets~\eqref{coset} with $P'<(N+1)P$~\cite{SW97}
without the use of the Bailey flow. 
 
The plan of the remainder of this paper is as follows. In sec.~\ref{sec_BF} we
summarize the results on Bailey flow which are needed for our construction.
In sec.~\ref{sec_bose} we construct the required Bailey pairs from the bosonic
polynomial generalizations given in~\cite{ABF,FB} of the characters 
$\chi_{r,s}^{(p,p')}$ of the minimal model $M(p,p')$ leaving the fermionic
side still undetermined. We establish the Bailey 
flow $M(p,p')\rightarrow (p',p'+N(kp'+p))_N$ for $k$ a nonnegative integer by
comparing the results of the bosonic side of the Bailey flow with the 
branching functions of the coset model~\eqref{coset} which are recalled 
in the appendix.
To make the Bose--Fermi identities explicit we need to write out the fermionic 
polynomials of the $M(p,p')$ models and here we find it convenient to consider 
$p'<2p$. Since in~\cite{BMS96} only the case $p'>2p$ was given in 
detail, we give the explicit fermionic side of the Bailey pairs in the desired 
regime in sec.~\ref{sec_fermi}. We present our final explicit results for the 
Bose--Fermi (or Rogers--Ramanujan) identities for the coset models $(P,P')_N$ 
in sec.~\ref{sec_identities} where the case $P'<(N+1)P$ is treated in 
sec.~\ref{sec_small} and $P'>(N+1)P$ in sec.~\ref{sec_big}. The fermionic form 
for the branching functions involves the same ``fractional-level'' Cartan 
matrix which arises in the study of the characters of 
$M(p,p')$~\cite{BM96,BMS96}. We conclude in sec.~\ref{sec_discussion} with a 
discussion of the relation between Bailey  and RG flow.

\section{The theory of Bailey flow}\label{sec_BF} 

In this section we summarize Bailey's original 
lemma~\cite{Bailey,Andrews84, Bressoud} and the recent results on the
higher-level Bailey lemma~\cite{SW96hl1,SW96hl2} which we will use.

\subsection{Bailey's lemma} 

Consider two sequences $\alpha=\{\alpha_L\}_{L\geq 0}$ and 
$\beta=\{\beta_L\}_{L\geq 0}$ which satisfy the relation
\begin{equation}\label{ab}
\beta_L = \sum_{i=0}^L \frac{\alpha_i}{(q)_{L-i}(aq)_{L+i}} .
\end{equation}
A pair $(\alpha,\beta)$ satisfying \eqref{ab} is called a Bailey pair
relative to $a$. 

Consider a second pair of sequences  
$\gamma=\{\gamma_L\}_{L\geq 0}$ and 
$\delta=\{\delta_L\}_{L\geq 0}$ which satisfies the relation
\begin{equation}\label{gd}
\gamma_L = \sum_{i=L}^{\infty} \frac{\delta_i}{(q)_{i-L}(aq)_{L+i}} \, ,
\end{equation}
which we call a conjugate Bailey pair. Bailey's lemma states that given a 
Bailey pair $(\alpha,\beta)$ and a conjugate Bailey pair $(\gamma,\delta)$, 
the following equation holds
\begin{equation}\label{obs}
\sum_{L=0}^{\infty}\alpha_L \gamma_L = \sum_{L=0}^{\infty}\beta_L \delta_L.
\end{equation}

In ref.~\cite{Bailey}, Bailey proved that the following $(\gamma,\delta)$ pair
satisfies~\eqref{gd}
\begin{align}\label{expgd}
\gamma_L&=\frac{(\rho_1)_L(\rho_2)_L(aq/\rho_1\rho_2)^L}
{(aq/\rho_1)_L(aq/\rho_2)_L(q)_{M-L}(aq)_{M+L}},\\[2mm]
\delta_L&=\frac{(\rho_1)_L(\rho_2)_L
(aq/\rho_1\rho_2)^L (aq/\rho_1\rho_2)_{M-L}
}{(aq/\rho_1)_M(aq/\rho_2)_M (q)_{M-L}}.
\end{align}
Using this pair in~\eqref{obs} yields
\begin{equation}\label{B_lemma}
\sum_{L=0}^M \frac{(\rho_1)_L (\rho_2)_L (aq/\rho_1\rho_2)^L}
{(aq/\rho_1)_L (aq/\rho_2)_L} \frac{\alpha_L}{(q)_{M-L} (aq)_{M+L}}
=\sum_{L=0}^M \frac{(\rho_1)_L (\rho_2)_L (aq/\rho_1\rho_2)^L
 (aq/\rho_1\rho_2)_{M-L}}{(aq/\rho_1)_M (aq/\rho_2)_M (q)_{M-L}}\beta_L,
\end{equation}
where the variables $\rho_1$ and $\rho_2$ can be chosen freely.
When $\rho_1,\rho_2 \to \infty$, ~\eqref{B_lemma}~ simplifies to
\begin{equation}\label{B_lemma_nr}
\sum_{L=0}^M \frac{a^L q^{L^2}}{(q)_{M-L} (aq)_{M+L}} \, \alpha_L
=\sum_{L=0}^M \frac{a^L q^{L^2}}{(q)_{M-L}} \, \beta_L,
\end{equation}
where we used
\begin{equation}
\lim_{\rho\to \infty} \rho^{-L}(\rho)_L=(-1)^L q^{\frac{L(L-1)}{2}}. 
\end{equation}

For later use we will need several results of Andrews~\cite{Andrews84} 
concerning the Bailey pair $(\alpha,\beta)$.

\subsubsection*{Dual Bailey pairs}
Given a Bailey pair $(\alpha,\beta)=(\alpha(a,q),\beta(a,q))$ relative to $a$, 
we may replace $q$ by $1/q$ to find that $(A,B)$ defined by
\begin{equation} \label{eqdual}
A_L = a^L q^{L^2} \, \alpha_L(a^{-1},q^{-1})
\qquad {\rm and} \qquad
B_L =  a^{-L} q^{-L(L+1)} \, \beta_L(a^{-1},q^{-1})
\end{equation}
is again a Bailey pair relative to $a$.
The pair $(A,B)$ is called the dual Bailey pair of $(\alpha,\beta)$.

\subsubsection*{Iterated Bailey pairs}
Let $(\alpha,\beta)$ be a Bailey pair relative to $a$. Then 
from~\eqref{B_lemma} the sequences $(\alpha',\beta')$ defined by
\begin{align} \label{a'}
\alpha'_L&= \frac{(\rho_1)_L (\rho_2)_L (aq/\rho_1\rho_2)^L}
{(aq/\rho_1)_L (aq/\rho_2)_L} \alpha_L \\  
\intertext{and} \label{b'}
\beta'_L&=\sum_{r=0}^L \frac{(\rho_1)_r (\rho_2)_r (aq/\rho_1\rho_2)^r
 (aq/\rho_1\rho_2)_{L-r}}{(aq/\rho_1)_L (aq/\rho_2)_L (q)_{L-r}} \beta_r
\end{align}
form again a Bailey pair relative to $a$. 
Equations (\ref{a'}) and (\ref{b'}) can of course be repeated an arbitrary 
number of times, leading to the Bailey chain
\begin{equation} \label{chain}
(\alpha,\beta)\to (\alpha',\beta')\to (\alpha'',\beta'')\to \cdots .
\end{equation}
Iterating \eqref{a'} and \eqref{b'} $k$ times with $\rho_1,\rho_2\to\infty$ 
yields the Bailey pair 
\begin{equation}\label{bchain}
\begin{split}
\alpha_L^{(k)}&=a^{kL}q^{kL^2}\alpha_L, \\
\beta_L^{(k)}&=\sum_{L\geq r_1\geq \ldots\geq r_k\geq 0}
\frac{a^{r_1+\cdots+r_k}q^{r_1^2+\cdots+r_k^2}}{(q)_{L-r_1}
(q)_{r_1-r_2}\ldots (q)_{r_{k-1}-r_k}} \: \beta_{r_k},
\end{split}
\end{equation}
where $k\geq 1$.

\subsection{Higher-level Bailey lemma}\label{sec_HLBL} 

In this paper we will use the new conjugate Bailey pairs found in  
refs.~\cite{SW96hl1,SW96hl2}. To state the result
needed, we denote by $\I$ the incidence matrix of the Lie algebra 
A$_{N-1}$, $\I_{j,k}=\delta_{j,k-1}+\delta_{j,k+1}$, and by $C$ the 
Cartan matrix $C=2I-\I$ with $I$ the identity matrix. The vectors
$\ve_i$ are the unit vectors $(\ve_i)_j=\delta_{i,j}$ for $i=1,\ldots,N-1$
and $\ve_i={\bs 0}$ for $i\neq 1,\ldots,N-1$.
Also define the $q$-binomial coefficient as
\begin{equation}
\label{qbin}
\Mult{m+n} {n}=\begin{cases} \cfrac{(q)_{m+n}}{(q)_m (q)_n} & 
\text{for } m,n\in \Integer_+,\\ 
0& \text{otherwise}, \end{cases}
\end{equation}
where $\Integer_+$ denotes the set of non-negative integers.

Then for $M\geq 0$, $N\geq 1$, $\lambda\geq 0$, $0\leq \ell <N$ 
and $\sigma=0,1$, the following conjugate Bailey pair 
relative to $a=q^{\lambda}$ is given in corollary 2.1 of ref.~\cite{SW96hl2}
\begin{equation}\label{gd_hl}
\begin{split}
\gamma_L&=\cfrac{a^{L/N}q^{L^2/N}}{(q)_{M-L}(aq)_{M+L}}
\srs{\vet}{L+(\lambda-\ell)/2}{\sigma}
q^{\vet C^{-1}(\vet-\ve_{\ell})} \prod_{j=1}^{N-1}
\Mult{\mu_j+\eta_j}{\eta_j},\\[1mm]
\delta_L&=\cfrac{a^{L/N}q^{L^2/N}}{(q)_{M-L}}
\srs{\vn}{L+(\lambda-\ell)/2}{\sigma}
q^{\vn C^{-1} (\vn-\ve_{\ell})} \prod_{j=1}^{N-1}
\Mult{m_j+n_j}{n_j},
\end{split}
\end{equation}
where $\vmu,\vet$ and $\vm,\vn$ are related by
\begin{align}
\label{et_mu}
\vmu+\vet &= \frac{1}{2}\Bigl(\I\,\vmu +(M-L)\,\ve_1
+(M+L+\lambda)\,\ve_{N-1}+\ve_{\ell}\Bigr),\\
\label{mn}
\vm+\vn &= \frac{1}{2} \Bigl(\I\,\vm+(2L+\lambda)\,\ve_{N-1}+\ve_{\ell}\Bigr).
\end{align}
The sums of the type $$\srs{\vn}{m}{\sigma}$$ are taken over the vector 
$\vn\in\Integer_+^{N-1}$ such that $\frac{m}{N}-(C^{-1}\vn)_1$ is an integer 
when $\sigma=0$ and half an odd integer when $\sigma=1$.

Before using the above conjugate Bailey pair to derive the
higher-level Bailey lemma, we make some observations.
Since the $q$-binomials are defined to be zero when its entries are
non-integer or negative we only need to sum over those 
$\vet,\vn\in \Integer_+^{N-1}$ 
in \eqref{gd_hl} which yield $\vmu,\vm \in \Integer_+^{N-1}$ 
from \eqref{et_mu} and \eqref{mn}. Using further 
$C^{-1}_{i,j}=\min\{i,j\}-ij/N$ this 
implies that $(L-(\lambda-\ell)/2)/N-C^{-1} \vv$ ($\vv=\vet,\vn$) is half 
an integer, explaining the restrictions on the above sums. The fact that one 
can in fact independently choose this expression to be half an even integer 
$(\sigma=0)$ or half an odd
integer $(\sigma=1)$, follows from other considerations~\cite{SW96hl2}.
We also note that since $C^{-1} \vv$ is a multiple of $1/N$, we only
obtain a non-trivial (nonzero) conjugate Bailey pair if we take
$\lambda+\ell+N\sigma$ to be even.

We may now insert the conjugate Bailey pair~\eqref{gd_hl} into~\eqref{obs}.
For later use we take $M\to\infty$ and eliminate $\vn$ in favour of $\vm$ 
via~\eqref{mn}. This gives the following lemma:
\begin{lemma}[higher-level Bailey lemma]\label{hlblem}
Fix integers $N\geq 1$, $\lambda\geq 0$, $0\leq \ell <N$ and $\sigma=0,1$,
such that $\ell+\lambda+N\sigma$ is even, and let
$(\alpha,\beta)$ form a Bailey pair relative to $q^{\lambda}$. 
Then the following identity holds:
\begin{multline}\label{lemma}
\frac{1}{(q)_{\infty}}
\sum_{L=0}^{\infty} q^{L(L+\lambda)/N} \, \alpha_L
\srs{\vet}{L+(\lambda-\ell)/2}{\sigma}
\frac{q^{\vet \, C^{-1} (\vet-\ve_{\ell})}}
{(q)_{\eta_1} \ldots (q)_{\eta_{N-1}}} \\
= \frac{q^{-\frac{\ell(N-\ell)+\lambda^2}{4N}}}{(q)_{\lambda}} 
\sum_{L=0}^{\infty} q^{\frac{1}{4}(2L+\lambda)^2} \beta_L
\sum_{\substack{\vm\in\Integer_+^{N-1}\\ \vm\equiv \vQ \mod{2}}}
q^{\frac{1}{4}\vm C\vm-\frac{1}{2}(2L+\lambda)m_{N-1}}
\prod_{j=1}^{N-1} \Mult{m_j+n_j}{m_j},
\end{multline}
where $\vm\equiv \vQ \pmod{2}$ stands for $m_j$ even when $Q_j$ is even and
$m_j$ odd when $Q_j$ is odd, and 
\begin{equation}
\vQ=(\ve_{\ell+1}+\ve_{\ell+3}+\cdots)+(\sigma+1)(\ve_1+\ve_3+\cdots).
\end{equation}
\end{lemma}
Note that for $N=1$ lemma \ref{hlblem} reduces to~\eqref{B_lemma_nr} with
$M\to\infty$ and $a=q^{\lambda}$.

\section{Bosonic polynomials for the minimal models $M(p,p')$ and 
Bailey flow}\label{sec_bose} 

\subsection{Bosonic polynomials} 

In order to make effective use of any of the forms of Bailey's lemma
given in the preceeding section it is necessary to find solutions of
the defining equation~\eqref{ab} for Bailey pairs. Here we derive such pairs 
from polynomial identities for finitizations of the characters of the minimal 
models $M(p,p')$ given in~\cite{BMS96},
\begin{equation}\label{bfpoly}
B^{(p,p')}_{r,s}(L,b)=F^{(p,p')}_{r,s}(L,b),
\end{equation}
where $1\leq b,s\leq p'-1$, $1\leq r\leq p-1$.
The bosonic function $B_{r,s}^{(p,p')}(L,b)$ is given by
\begin{multline}\label{bose}
B^{(p,p')}_{r,s}(L,b)\\
=\sum_{j=-\infty}^{\infty} \left(
q^{j(jpp'+rp'-sp)} \Mult{L}{\frac{1}{2}(L+s-b)-jp'}
-q^{(jp+r)(jp'+s)}\Mult{L}{\frac{1}{2}(L-s-b)-jp'} \right),
\end{multline}
which first appeared in the work of Andrews, Baxter and Forrester~\cite{ABF} 
for $p'=p+1$, and for general $p,p'$ in the work of Forrester and 
Baxter~\cite{FB}.
Since the Bailey flows can be determined without knowledge of the explicit
form of the fermionic side, we postpone giving $F_{r,s}^{(p,p')}(L,b)$
until section~\ref{sec_fermi}.

The limit $L\to \infty$ of the polynomials~\eqref{bose} are 
the characters of the minimal models $M(p,p')$, namely
$\lim_{L\to \infty}B_{r,s}^{(p,p')}(L,b)=\hat{\chi}_{r,s}^{(p,p')}(q)$
where~\cite{FF83,RC83,CIZ87}
\begin{equation}
\chi^{(p,p')}_{r,s}(q)=q^{\Delta_{r,s}-\frac{c}{24}} 
\hat{\chi}^{(p,p')}_{r,s}(q),
\end{equation}
with 
\begin{equation} \label{RC}
\hat{\chi}^{(p,p')}_{r,s}(q)=
\hat{\chi}^{(p,p')}_{p-r,p'-s}(q)=\frac{1}{(q)_{\infty}}
\sum_{j=-\infty}^{\infty} \left( q^{j(jpp'+rp'-sp)}- 
q^{(jp+r)(jp'+s)} \right).
\end{equation}
The central charge and conformal dimensions are given by
\begin{equation}
c=1-\frac{6(p'-p)^2}{pp'}\qquad \text{and} \qquad
\Delta_{r,s}=\frac{(rp'-sp)^2-(p'-p)^2}{4pp'},
\end{equation}
respectively.  

\subsection{Bailey pairs}\label{secBP} 

To obtain Bailey pairs from~\eqref{bfpoly}, we use the definition of the 
$q$-binomials~\eqref{qbin} in~\eqref{bfpoly} and replace 
$L$ by $2L+\lambda$ where $\lambda=|b-s|$. Then multiplying by 
$(q)_{\lambda}/(q)_{2L+\lambda}$
we note that the resulting left-hand side is in the form of the
right-hand side of~\eqref{ab} and we find the following Bailey pairs 
relative to $q^{\lambda}=q^{|b-s|}$.

\subsubsection*{Bailey pair arising from \eqref{bfpoly} and \eqref{bose}}
\begin{equation}\label{BP}
\begin{split}
\alpha_L & = \begin{cases}
1 & \text{ for } L=0, \\
q^{j(jpp'\pm rp' \mp sp)}&\text{ for } 
L=jp'-(\lambda \mp b \pm s)/2,~(j\geq 1),\\
-q^{(jp+r)(jp'+s)}&\text{ for } L=jp'-(\lambda-b-s)/2,~(j\geq 0),\\
-q^{(jp-r)(jp'-s)}&\text{ for } L=jp'-(\lambda+b+s)/2,~(j\geq 1), \\
0 & \text{ otherwise},
\end{cases} \\
\beta_L&=
\frac{(q)_{\lambda}}{(q)_{2L+\lambda}} F^{(p,p')}_{r,s}(2L+\lambda,b).
\end{split}
\end{equation}

\subsubsection*{Dual Bailey pair arising from \eqref{bfpoly} and \eqref{bose}}
\begin{equation}\label{dBP}
\begin{split}
\alpha_L&= \begin{cases}
1 & \text{ for } L=0, \\
q^{j(j(p'-p)p'\pm (b-r)p'\mp s(p'-p))}&
\text{ for } L=jp'-(\lambda \mp b \pm s)/2,~(j\geq 1),\\
-q^{(j(p'-p)+b-r)(jp'+s)}&\text{ for } L=jp'-(\lambda-b-s)/2,~(j\geq 0),\\
-q^{(j(p'-p)-b+r)(jp'-s)}&\text{ for } L=jp'-(\lambda+b+s)/2,~(j\geq 1), \\
0 & \text{ otherwise},
\end{cases} \\
\beta_L&= q^{L(L+\lambda)}\frac{(q)_{\lambda}}{(q)_{2L+\lambda}} 
F^{(p,p')}_{r,s}(2L+\lambda,b;1/q).
\end{split}
\end{equation}

\subsubsection*{Iterated Bailey pair arising from \eqref{bfpoly} and
\eqref{bose}}
\begin{equation}\label{BPchain}
\begin{split}
\alpha^{(k)}_L&=
\begin{cases}
1 & \text{ for } L=0, \\
q^{j(jp'(p+kp')\pm p'(r+kb)\mp s(p+kp'))}&
\text{ for } L=jp'-(\lambda\mp b\pm s)/2,~(j\geq 1)\\
-q^{(jp'+s)(j(p+kp')+kb+r)}&\text{ for }L=jp'-(\lambda-b-s)/2,~(j\geq 0),\\
-q^{(jp'-s)(j(p+kp')-kb-r)}&\text{ for }L=jp'-(\lambda+b+s)/2,~(j\geq 1), \\
0 & \text{ otherwise},
\end{cases}\\[2mm]
\beta^{(k)}_L&=
\sum_{L\geq r_1\geq \ldots\geq r_k\geq 0}
\frac{q^{r_1(r_1+\lambda)+\cdots+r_k(r_k+\lambda)}}{(q)_{L-r_1} (q)_{r_1-r_2}
\cdots (q)_{r_{k-1}-r_k}}\frac{(q)_{\lambda}}{(q)_{2r_k+\lambda}}
F_{r,s}^{(p,p')}(2r_k+\lambda,b).
\end{split}
\end{equation}
 
\subsubsection*{Iterated dual Bailey pair arising from \eqref{bfpoly} and
\eqref{bose}}
\begin{equation}\label{dBPchain}
\begin{split}
\alpha^{(k)}_L&=
\begin{cases}
1 & \text{ for } L=0, \\
q^{j(jp'(kp'+p'-p)\pm p'(kb+b-r)\mp s(kp'+p'-p))}&
\text{ for }L=jp'-(\lambda\mp b \pm s)/2,~(j\geq 1),\\
-q^{(jp'-s)(j(kp'+p'-p)-kb-b+r)}&
\text{ for }L=jp'-(\lambda-b-s)/2,~(j\geq 0),\\
-q^{(jp'+s)(j(kp'+p'-p)+kb+b-r)}&
\text{ for }L=jp'-(\lambda+b+s)/2,~(j\geq 1), \\
0 & \text{ otherwise},
\end{cases}\\[2mm]
\beta^{(k)}_L&=
\sum_{L\geq r_1\geq \ldots \geq r_{k}\geq 0}
\frac{q^{r_1(r_1+\lambda)+\cdots+r_{k-1}(r_{k-1}+\lambda)+2r_k(r_k+\lambda)}}
{(q)_{L-r_1} (q)_{r_1-r_2}\cdots (q)_{r_{k-1}-r_{k}}} 
\frac{(q)_{\lambda}}{(q)_{2r_k+\lambda}}F_{r,s}^{(p,p')}(2r_{k}+\lambda,b;1/q).
\end{split}
\end{equation}

\subsection{Bailey flow}\label{sec_Bflow} 

We use the explicit expressions for $\alpha_L$ for the four sets
of Bailey pairs derived previously to establish a Bailey flow
from the minimal models $M(p,p')$ to the A$^{(1)}_1$ cosets.
The normalized branching functions of these cosets, obtained from~\cite{ACT} 
and denoted $\hat{\chi}_{r,s;\ell}^{(P,P';N)}(q)$,
are given in the appendix in equation~\eqref{bfN} for $N$ integer and $N'$ 
fractional.

Substituting $\alpha_L$ of equations~\eqref{BP} and \eqref{BPchain}
into the left-hand side of the higher-level Bailey lemma~\eqref{lemma} yields,
after using the symmetry properties~\eqref{sfsym},
\begin{equation}
\hat{\chi}^{(p',p'+N(p+kp');N)}_{s,b+N(kb+r);\ell}(q)
\end{equation}
for $k\geq 0$ and $\ell+\lambda+N(kb+r)$ even.

Similarly, from $\alpha_L$ in equations~\eqref{dBP} and \eqref{dBPchain}
we find that the left-hand side of~\eqref{lemma} becomes
\begin{equation}
\hat{\chi}^{(p',p'+N(kp'+p'-p);N)}_{s,b+N(kb+b-r);\ell}(q)
\end{equation}
with $k\geq 0$ and $\ell+\lambda+N(kb+b-r)$ even.

Hence we have demonstrated the following Bailey flows
\begin{align}\label{hl_flow}
M(p,p')=(p,p')_1 &\to \left(p',p'+N(kp'+p)\right)_N \quad &(k\geq 0)\\
\intertext{and}
\label{dhl_flow}
M(p,p')=(p,p')_1 &\dto \left(p',p'+N(kp'+p'-p)\right)_N \quad &(k\geq 0).
\end{align}
The flows~\eqref{hl_flow} and~\eqref{dhl_flow} show that the spectra of 
the cosets~\eqref{coset} with $N\geq 2$ can be expressed entirely in terms 
of truncated or finitized spectra of the $c<1$ theories, a property noted 
to hold for the unitary models by Nakayashiki and Yamada~\cite{NY}.

We note that if we start with $M(p'-p,p')$ instead of $M(p,p')$ the
Bailey flow and dual Bailey flow interchange
\begin{align}\label{duality}
M(p'-p,p') &\dto \left(p',p'+N(kp'+p)\right)_N, \notag\\
M(p'-p,p') &\to \left(p',p'+N(kp'+p'-p)\right)_N.
\end{align}
Thus we see that to obtain fermionic representations for the $(P,P')_N$
theories using the right-hand side of the higher level Bailey lemma,
it is sufficient to restrict our attention to $M(p,p')$ with $p<p'<2p$.

\section{The fermionic polynomials for $M(p,p')$ with $p<p'<2p$}
\label{sec_fermi} 
In this section we give the explicit forms for 
$F_{r,s}^{(p,p')}(L,b)$ in identity~\eqref{bfpoly} for $p<p'<2p$. 
For the special cases $M(2k-1,2k+1)$ and $M(p,p+1)$ the fermionic forms were 
proven in \cite{FQ95,Kir} and~\cite{B94}-\cite{Ole1,Ole2},
respectively. For general $p,p'$ the results were proven in~\cite{BM96,BMS96}.
Since in~\cite{BMS96} explicit formulas for $F_{r,s}^{(p,p')}(L,b)$ were 
given for $p'>2p$ only, we state the formulas for the dual regime $p<p'<2p$ 
here, which follow from the formulas for $p'>2p$ by taking $q\to 1/q$.

In \eqref{bfpoly} we implicitly gave the Bose--Fermi polynomial
identities for the minimal models for all $1\leq b,s\leq p'-1$ and
$1\leq r \leq p-1$. However, the explicit form for $F_{r,s}^{(p,p')}(L,b)$
is so far only known for several special cases. In particular we cannot
treat the variables $b$ and $r$ independently, and the choice of $b$
fixes $r$. As a further restriction only certain values of $s$ and
$b$ have been treated at present. Hence in the remainder of this paper
we will only deal with a subset of all possible $F_{r,s}^{(p,p')}(L,b)$. 

The fermionic functions $F_{r,s}^{(p,p')}(L,b)$ are much more involved than the
bosonic functions \eqref{bose} and several preliminary sections are needed
to introduce all the necessary notations and definitions.
In sec.~\ref{sec_TD} we review the continued fraction expansion of 
$p'/(p'-p)$ and the closely related Takahashi--Suzuki decomposition.
Then, in sec.~\ref{sec_fff}, we define fundamental fermionic functions which 
are the building blocks of $F_{r,s}^{(p,p')}(L,b)$. In sec.~\ref{sec_eff},
$F_{r,s}^{(p,p')}(L,b)$ is given for four classes of values of $b$
(and hence $r$), listed in table~\ref{tab}. 

\subsection{Takahashi--Suzuki decomposition}\label{sec_TD} 
Given $p$ and $p'$ such that $\gcd(p,p')=1$ and such that $p<p'<2p$,
we define integers $\nu_0,\dots,\nu_n$ by the continued fraction expansion
\begin{equation}\label{cfrac}
\frac{p'}{p'-p}=1+ \nu_0+\cfrac{1}{\nu_1+\cfrac{1}{\nu_2+\cdots
+\cfrac{1}{\nu_n+2}}} \: .
\end{equation}
The number $n+1$ is referred to as the number of zones, $\nu_i$ being the
size of the $i$-th zone.
Using the $\nu_i$'s we define another set of integers $t_1,\dots,t_{n+1}$ as
\begin{equation}\label{deft}
t_i=\sum_{j=0}^{i-1}\nu_j.
\end{equation}
For convenience we also set $t_0=-1$.
Given these integers we define a generalized or fractional-level 
incidence matrix ${\cal I}_B$ with entries
\begin{equation}\label{I_B}
({\cal I}_B)_{j,k} = \begin{cases}
\delta_{j,k+1} + \delta_{j,k-1} & \text{for 
$1 \leq j<t_{n+1},~ j \neq t_i$}, \\
\delta_{j,k+1} + \delta_{j,k} - \delta_{j,k-1} & \text{for 
$j=t_i,~1\leq i\leq n-\delta_{\nu_n,0}$}, \\
\delta_{j,k+1} + \delta_{\nu_n,0} \delta_{j,k} & \text{for $j=t_{n+1}$},
\end{cases}
\end{equation}
and a fractional-level Cartan matrix $B$ as
\begin{equation} \label{defB}
B=2 I -{\cal I}_B.
\end{equation}

To give the fermionic forms $F_{r,s}^{(p,p')}(L,b)$ we need to decompose
$r,s$ and $b$ in terms of Takahashi--Suzuki (TS) lenghts~\cite{TS}. 
To describe this decomposition we first define the recursions
\begin{align}\label{def_y}
y_{m+1}&=y_{m-1}+(\nu_m+\delta_{m,0}+2\delta_{m,n}) y_m, 
& y_{-1}&=0, & y_0&=1,\\
\label{def_yp}
\by_{m+1}&=\by_{m-1}+(\nu_m+\delta_{m,0}+2\delta_{m,n}) \by_m, & 
\by_{-1}&=-1, & \by_0&=1.
\end{align}
For $j=0,\dots,t_{n+1}+1$, the TS lengths $l_{j+1}$ and 
truncated TS lengths $\bl_{j+1}$ are then given by 
\begin{equation}\label{taka}
\begin{aligned}
l_{j+1}&=y_{m-1}+(j-t_m)y_m\\
\bl_{j+1}&=\by_{m-1}+(j-t_m)\by_m\\
\end{aligned}
\qquad \text{ for } t_m<j\leq t_{m+1}+\delta_{m,n} \text{ with } 
0\leq m\leq n. 
\end{equation}
An arbitrary integer $b$ ($1\leq b<p'$) 
may be uniquely decomposed into TS lengths as
\begin{equation}\label{t_decom}
b=\sum_{i=1}^{k} l_{\beta_i+1}
\end{equation}
provided that $t_{\xi_i}<\beta_i \leq t_{\xi_i+1}+\delta_{\xi_i,n}$ with
$0\leq \xi_1<\xi_2<\cdots <\xi_k\leq n$ with the additional restriction, 
$\xi_{i+1}\geq \xi_i+2$ when $\beta_i=t_{\xi_i+1}$.

Finally, $t_{n+1}$-dimensional vectors $\vQ^{(j)}$ 
$(j=1,\ldots,t_{n+1}+1)$ are needed to
specify parities of summation variables.
For $1\leq i\leq t_{n+1}$ and $0\leq m\leq n$ such that 
$t_m<j\leq t_{m+1}+\delta_{m,n}$ 
the components of $\vQ^{(j)}$ are recursively defined as
\begin{equation}\label{eqn_Qij}
Q^{(j)}_{i}=\begin{cases} 0 & \text{for $j \leq i \leq t_{n+1}$},\\  
j-i & \text{for $t_m\leq i<j$},\\
Q^{(j)}_{i+1}+Q^{(j)}_{t_{m'}+1} 
& \text{for $t_{m'-1}\leq i<t_{m'},~1\leq m'\leq m$}.
\end{cases}
\end{equation}
When $\nu_n=0$, so that $t_{n+1}=t_n$, we need to set the initial
condition $Q_{t_n+1}^{(t_n+1)}=0$.

\subsection{The fundamental fermionic functions}\label{sec_fff} 

With the definitions of the previous subsection we now introduce
fundamental fermionic functions $f$ and $\tilde f$ used as building blocks
for $F_{r,s}^{(p,p')}(L,b)$.
First, for $\vu,\vv\in\Integer^{t_{n+1}}$,
\begin{equation}\label{f_fermi}
f(L,\vu,\vv)=
\sum_{\substack{\vm\in\Integer_+^{t_{n+1}}\\ \vm\equiv\vQ_{\vu+\vv}\mod{2}}}
q^{\frac{1}{4}\vm B \vm -\frac{1}{2}\vA_{\vu,\vv}\vm}
\prod_{j=1}^{t_{n+1}}\Mult{n_j+m_j}{m_j}',
\end{equation}
with $B$ as in~\eqref{defB}, the $(\vm,\vn)$-system given by 
$\vm+\vn=\frac{1}{2}(\I_{B}\vm+\vu+\vv+L\ve_1)$,
$$ \vQ_{\vu}=\sum_{j=1}^{t_{n+1}+1} u_j \vQ^{(j)},$$
and, for $t_i<j\leq t_{i+1}$,
\begin{equation}\label{Ag}
(\vA_{\vu,\vv})_j= \begin{cases} 
u_j & \text{for $i$ odd},\\
v_j & \text{for $i$ even}. \end{cases}
\end{equation}
The notation $\vm\equiv \vQ \pmod{2}$ stands for $m_j$ even 
when $Q_j$ is even and
$m_j$ odd when $Q_j$ is odd.
The $q$-binomials $\Mult{n+m}{m}'$ in \eqref{f_fermi} differ slightly 
from those of~\eqref{qbin},
\begin{equation}\label{qbin_prime}
\Mult{n+m}{m}'= \begin{cases}
 \cfrac{(q^{n+1})_m}{(q)_m} & \text{for $m\in\Integer_+$, $n\in\Integer$},\\
 0 & \text{otherwise}.
\end{cases}
\end{equation}
Notice that for negative $n$, $\Mult{n+m}{m}'$ can only be nonzero  
when $n+m<0$.

The second function $\tf$ is defined
for the special vectors $\vu=\ve_{\nu_0-j-1}-\ve_{\nu_0}+\vu'$
with $0\leq j<\nu_0$ and $(\vu')_i=0$ for $1\leq i\leq \nu_0$,
\begin{equation}\label{f_tilde}
\tf(L,\vu,\vv) = \begin{cases}
q^{\frac{L}{2}}f(L,\vu,\vv)
+(1-q^L)f(L-1,\vu-\ve_{\nu_0-1}+\ve_{\nu_0},\vv) & \text{for $j=0$},\\[1mm]
q^{\frac{j}{2}}\bigl[f(L+1,\vu-\ve_{\nu_0-j-1}+\ve_{\nu_0-j},\vv) \\
\qquad 
-q^{\frac{L+1}{2}}f(L,\vu-\ve_{\nu_0-j-1}+\ve_{\nu_0-j+1},\vv) \bigr]
& \text{for $1\leq j<\nu_0$}.\end{cases}
\end{equation}

We make a final remark about the notation employed in this and subsequent
sections. In section~\ref{sec_HLBL} the vectors $\ve_j$ were defined
as $(N-1)$-dimensional unit vectors. In the above we clearly mean
$\ve_j$ to be $t_{n+1}$-dimensional. Indeed we will throughout use
$\ve_j$ for the $j$-th unit vector, and assume that its dimension is clear
from the context.

\subsection{Explicit fermionic functions}\label{sec_eff} 
Using the functions $f$ and $\tilde{f}$ as fundamental objects,
we now give explicit forms for $F_{r,s}^{(p,p')}(L,b)$,
with $p<p'<2p$ such that $\nu_0,\nu_n\geq 1$ and $\nu_j>1$ ($j=1,\ldots,n-1$).
We consider here $s$ being a single TS length
$$s=l_{\sigma+1} \text{ with } t_{\varsigma}<\sigma\leq t_{\varsigma+1}
+\delta_{\varsigma,n}$$
and we limit ourselves to $b$ being one of the cases listed in table~\ref{tab}.
Case 1 corresponds to $b$ being a single TS length and case 2 deals with
values of $b$ in the vicinity of a single TS length. Cases 3 and 4 are rather
generic and deal with classes of $b$ having a TS decomposition with TS 
lengths in adjacent zones (starting in zone 0 for case 3 and in a zone $>0$ 
for case 4).
The labels $\beta_m$ are restricted to lie not too close to the upper boundary
of the zone. The reason for having to distinguish between many different
cases (of which we have only listed a few) is that $F_{r,s}^{(p,p')}(L,b)$
is a linear combination of $f(L,\vu,\vv)$ and $\tf(L,\vu,\vv)$
where the vectors $\vu$ in this linear combination can be determined
by a recursive procedure~\cite{BMS96}. To give explicit formulas for the
vectors $\vu$ as arising from this recursive procedure one needs to 
distinguish many cases.

As mentioned before, the choice of $b$ fixes $r$ and given $b$ and its 
decomposition (\ref{t_decom}) in TS lengths, $r$ reads
\begin{equation}\label{r(b)}
r(b)= \begin{cases}
\sum_{i=1}^k \bl_{\beta_i+1} &
\text{for $1\leq b<p'-\nu_0$},\\
\sum_{i=1}^{\beta} \bl_{\beta_i+1}+1 &
\text{for $p'-\nu_0\leq b\leq p'-1$}.
\end{cases}
\end{equation}

\begin{table} 
\begin{equation*} 
\begin{array}{|l|l|l|l|}
\hline
\multicolumn{1}{|c|}{{\rm Case}} & \multicolumn{1}{c|}{b} & 
\multicolumn{1}{c|}{r} & \multicolumn{1}{c|}{{\rm additional~restrictions}}\\
\hline \hline 
1: & b=l_{\beta+1} & r=\bl_{\beta+1} & 
 \begin{aligned}[c] 
  & t_{\xi}<\beta\leq t_{\xi+1}+\delta_{\xi,n}\\
  & 0\leq \xi\leq n\\
 \end{aligned}
\\ \hline
2a: & b=l_{\beta+1}-j & r=\bl_{\beta+1} & 
 \begin{aligned}[c]
  &t_{\xi}+\delta_{\xi,1}<\beta <t_{\xi+1}+\delta_{\xi,n}\\
  &1\leq \xi\leq n,~~1\leq j< \nu_0
 \end{aligned}\\ 
\hline
2b: & b=l_{\beta+1}+1+j & 
 r=1+\bl_{\beta+1} & 
 \begin{aligned}[c]
  &t_{\xi}<\beta <t_{\xi+1}+\delta_{\xi,n}\\
  &1\leq \xi\leq n,~~0\leq j\leq \nu_0
 \end{aligned}\\
\hline
3: & b=\sum_{m=0}^{\xi} l_{\beta_m+1} & 
 r=\sum_{m=0}^{\xi} \bl_{\beta_m+1} & 
 \begin{aligned}[c]
  &0\leq \beta_0\leq \nu_0-1\\ 
 &t_m<\beta_m\leq t_{m+1}-3~~(1\leq m\leq \xi-2)\\
 &t_{\xi-1}< \beta_{\xi-1}\leq t_{\xi}-2\\
 &t_{\xi}<\beta_{\xi}\leq t_{\xi+1}-1\\
 &1\leq \xi\leq n
 \end{aligned}\\
\hline
4: & b=\sum_{m=\zeta}^{\xi} l_{\beta_m+1} 
& r=\sum_{m=\zeta}^{\xi} \bl_{\beta_m+1} & 
 \begin{aligned}[c]
  &t_m<\beta_m\leq t_{m+1}-3~~(\zeta\leq m\leq \xi-2)\\
  &t_{\xi-1}<\beta_{\xi-1}\leq t_{\xi}-2\\
  &t_{\xi}<\beta_{\xi}\leq t_{\xi+1}-1\\
  &1\leq \zeta<\xi\leq n
 \end{aligned}\\
\hline
\end{array}
\end{equation*}
\caption{\label{tab} Cases for $b$ considered}
\end{table}

In all four cases for $b$ an overall normalization constant
$k_{b,s}$ is fixed by the condition
$$ \left.F_{r,s}^{(p,p')}(L,b)\right|_{q=0}=1.  $$
Furthermore, in all of the cases below, we use the abbreviation $\vu_s$
for
$$\vu_s=\ve_{\sigma}-\vE_{\varsigma+1,n} \quad \text{where}\quad 
\vE_{i,j}=\sum_{k=i}^j \ve_{t_k}.$$

\subsubsection*{Case 1:}
\begin{equation} \label{case1}
F_{r,s}^{(p,p')}(L,b)=q^{k_{b,s}}f(L,\ve_{\beta}-\vE_{\xi+1,n},\vu_s).
\end{equation}

\subsubsection*{Case 2a:}
\begin{multline} \label{case2a}
F_{r,s}^{(p,p')}(L,b)=
q^{k_{b,s}}\Bigl\{ q^{\frac{-2\nu_0+\co{\xi}+2j}{4}}
\tf(L,\ve_{j-1}+\ve_{\beta-1}-\vE_{1,n},\vu_s)\\
+\sum_{i=2}^{\xi} q^{\frac{-2\nu_0+\ce{i}+2j}{4}}
\tf(L,\ve_{j-1}+\ve_{t_i-1}+\ve_{\beta}-\vE_{1,i}-\vE_{\xi+1,n},\vu_s)\\
+ f(L,\ve_{\nu_0-j}-\ve_{\nu_0}+\ve_{\beta}-\vE_{\xi+1,n},\vu_s) \Bigr\},
\end{multline}
with $\theta(\text{true})=1$ and $\theta(\text{false})=0$. 

\subsubsection*{Case 2b:}
\begin{multline} \label{case2b}
F_{r,s}^{(p,p')}(L,b)=
q^{k_{b,s}} \left\{ q^{\frac{\nu_0+1-3\co{\xi}}{4}}
f(L,\ve_{j}+\ve_{\beta+1}-\vE_{1,n},\vu_s)\right.\\
+\sum_{i=2}^{\xi} q^{\frac{\nu_0-\co{i}}{4}}
f(L,\ve_{j}+\ve_{t_i-1}+\ve_{\beta}-\vE_{1,i}-\vE_{\xi+1,n},\vu_s)\\
\left. +\theta(j<\nu_0)q^{\frac{\nu_0-2j-1}{4}} 
\tf(L,\ve_{\nu_0-j-1}-\ve_{\nu_0}+\ve_{\beta}-\vE_{\xi+1,n},\vu_s) \right\}.
\end{multline}

For the remaining two cases some more notation is needed.  Let
\begin{equation}\label{b_ab}
b=\sum_{m=\zeta}^{\xi}l_{\beta_m+1},\quad \text{ with }
t_m<\beta_m\leq t_{m+1}+\delta_{m,n} 
\text{ and } 0\leq \zeta\leq m\leq \xi \leq n.
\end{equation}
Then we define two vectors
\begin{align}
\vbeta&=(\beta_{\zeta},\beta_{\zeta+1},\ldots,\beta_{\xi-1},\beta_{\xi}),\\
\vi&=(i_{\zeta},i_{\zeta+1},\ldots,i_{\xi-1},0),
\end{align}
where the $\beta_m$ ($\zeta\leq m\leq \xi$) are defined from (\ref{b_ab}) and
$i_m\in\{0,1\}$ ($\zeta\leq m<\xi$). Since the components of $\vi$ are 
$0$ or $1$, the vector $\vi$ may also be represented by a binary word
\begin{equation}\label{i_ab}
\vi=1^{b_1}0^{a_1}1^{b_2}0^{a_2}\cdots 1^{b_{\ell}}0^{a_{\ell}},
\quad \text{with } 
\begin{cases}
a_k\geq 1,& 1\leq k\leq \ell,\\
b_k\geq 1,& 2\leq k\leq \ell,~~b_1\geq 0.
\end{cases}
\end{equation}
The numbers $a_k$ and $b_k$ give the length of the $k$-th substring 
of $0$'s and $1$'s (starting with a string
of 1's of possibly zero length).
Finally let us define for $t_m<j\leq t_{m+1}+\delta_{m,n}$
$$R(j+1) = t_{m+1}+t_m-j-1,$$
and in addition $R_0=Id$ and $R_1=R$.
This prepares us for the cases 3 and 4.

\subsubsection*{Case 3:}
\begin{align}\label{case3}
F_{r,s}^{(p,p')}(L,b)=
q^{k_{b,s}} & \Bigl\{ 
\sum_{\substack{i_1,\ldots,i_{\xi-1}=0,1\\ i_0=0}} 
q^{-\varphi_{\vi}} 
f(L,\ve_{\beta_0+i_1}-\ve_{\nu_0}+\vu_{\vi,\vbeta},\vu_s) \Bigr.\notag\\
& \Bigl. +\sum_{\substack{i_1,\ldots,i_{\xi-1}=0,1\\ i_0=1}}
q^{-\psi_{\vi}} 
\tf(L,\ve_{\nu_0-\beta_0-i_1}-\ve_{\nu_0}+ \vu_{\vi,\vbeta},\vu_s) \Bigr\},
\end{align}
where
\begin{align}
\vu_{\vi,\vbeta}
&=\sum_{m=1}^{\xi-1} 
\ve_{R_{i_m}(\beta_m+1)+|i_{m+1}-i_m|-|i_m-i_{m-1}|}
+\ve_{1+\beta_{\xi}-i_{\xi-1}} -\vE_{2,n}\notag \\
\varphi_{\vi}
&=\frac{1}{2}\sum_{j=2}^{\ell}
(-1)^{a_1+\sum_{k=2}^{j-1}(a_k+b_k)} \ce{b_j}\notag\\
\psi_{\vi}
&=\frac{\beta_0+i_1-1}{2}
+\frac{1}{2}\sum_{j=1}^{\ell} (-1)^{\sum_{k=1}^{j-1}(a_k+b_k)} \ce{b_j}
\end{align}
and $a_k,~b_k$ and $\ell$ follow from the representation \eqref{i_ab} of
$\vi$.

\subsubsection*{Case 4:}
\begin{multline}\label{case4}
F_{r,s}^{(p,p')}(L,b)=q^{k_{b,s}}
\Bigl[ \sum_{\substack{i_{\zeta+1},\ldots,i_{\xi-1}=0,1\\ i_{\zeta}=0}}
q^{-\varphi_{\vi}} f(L,\vu_{\vi,\vbeta}^{(1)})  \Bigr.\\  
+\Bigl. \sum_{\substack{i_{\zeta+1},\ldots,i_{\xi-1}=0,1\\ i_{\zeta}=1}}
\Bigl\{ q^{-\psi_{\vi}} 
f(L,\vu_{\vi,\vbeta}^{(2)},\vu_s) + 
q^{-\chi_{\vi}} f(L,\vu_{\vi,\vbeta}^{(3)},\vu_s) \Bigr\} \Bigr],
\end{multline}
where 
\begin{align}
\vu_{\vi,\vbeta}^{(1)}&=\ve_{\beta_{\zeta}+i_{\zeta+1}}
+\vu_{\vi,\vbeta}, \notag \\
\vu_{\vi,\vbeta}^{(2)}&=-\ve_{t_{\zeta}}+\ve_{t_{\zeta+1}
+t_{\zeta}-\beta_{\zeta} -i_{\zeta+1}-1}+\vu_{\vi,\vbeta}, \notag \\
\vu_{\vi,\vbeta}^{(3)}&=\ve_{t_{\zeta}-1}-\ve_{t_{\zeta}}+\ve_{t_{\zeta+1}
+t_{\zeta}-\beta_{\zeta}-i_{\zeta+1}}+\vu_{\vi,\vbeta}, \notag \\
\vu_{\vi,\vbeta}&=\sum_{m=\zeta+1}^{\xi-1} \ve_{R_{i_m}(\beta_m+1)
+|i_{m+1}-i_m|-|i_m-i_{m-1}|}+\ve_{\beta_{\xi}+1-i_{\xi-1}}-\vE_{\zeta+1,n}
\end{align}
and 
\begin{align}
\varphi_{\vi}&=\frac{1}{2}(-1)^{\zeta}\sum_{j=2}^{\ell} 
 (-1)^{a_1+\sum_{k=2}^{j-1}(a_k+b_k)} \ce{b_j},\notag\\
\psi_{\vi}&=(-1)^{\zeta} \Bigl( \frac{1}{4}(-1)^{b_1}+\frac{1}{2}
\sum_{j=2}^{\ell} (-1)^{\sum_{k=1}^{j-1}(a_k+b_k)}\ce{b_j}\Bigr),\notag\\
\chi_{\vi}&=-\frac{1}{2}(-1)^{\zeta}\Bigl( \co{b_1}+\sum_{j=2}^{\ell}
(-1)^{\sum_{k=1}^{j-1}(a_k+b_k)} \ce{b_j}\Bigr),
\end{align}
with $a_k,~b_k$ and $\ell$ as defined in \eqref{i_ab}.

\section{The explicit Bose--Fermi identities for the A$_1^{(1)}$ cosets}
\label{sec_identities} 
We are finally in the position to give explicit Bose--Fermi 
identities for the A$_1^{(1)}$ cosets~\eqref{coset}. 
Corresponding to the four classes of Bailey pairs given in
section~\ref{secBP} we treat the cases
(i) $(N/2+1)P<P'<(N+1)P$, (ii) $P<P'<(N/2+1)P$,
(iii) $(Nk+N/2+1)P<P'<(Nk+N+1)P$ and (iv) $(Nk+1)P<P'<(Nk+N/2+1)P$
with $k\geq 1$ separately.

A principle result of this section will be that in all four cases 
the fermionic branching functions for $(P,P')_N$ involve an
$(\vmh,\vnh)$-system determined from the continued fraction expansion
\begin{equation}\label{CFD}
\frac{N P'}{P'-P}=1+\nuh_0+\cfrac{1}{\nuh_1+\cfrac{1}{\nuh_2+\cdots
+\cfrac{1}{\nuh_{\nh}+2}}}\: 
\end{equation}
as observed for the cases (i) and (ii) in~\cite{SW97}.
In order to present the results of this section as compactly as possible we 
introduce a notation which distinguishes between variables that occur in the
$M(p,p')$ polynomials --the input in the Bailey lemma--
and the corresponding quantities for the branching functions of $(P,P')_N$.
To this end we adopt the notation that all quantities in the $(P,P')_N$ 
branching functions which have a counterpart $x$ for $M(p,p')$ will be denoted
by $\hat{x}$.
In particular, using the continued fraction expansion \eqref{CFD} we define 
$\th_i,\I_{\Bh},\Bh,\yh_m,\yhb_m,
\lh_{j+1},\hat{\bar{l}}_{j+1}$ and $\vQh^{(j)}$ 
by taking the corresponding definitions for $t_i, \I_B, B, y_m,
\by_m, l_{j+1}, \bl_{j+1}$ and ${\bf Q}^{(j)}$ 
of  section~\ref{sec_TD} and by replacing all variables therein by their 
hatted counterparts.

\subsection{Fermionic Branching Functions for $(P,P')_N$ with $P'<(N+1)P$}
\label{sec_small} 

As in section~\ref{sec_fermi}, we define a fundamental fermionic function
$\fh$ for the A$_1^{(1)}$ cosets~\eqref{coset}, which will 
be the building block of the fermionic branching functions.
For $\vuh,\vAh,\vQh\in\Integer^{\th_{\nh+1}}$ we set 
\begin{equation}\label{f_hat}
\fh(\vuh,\vAh,\vQh)=
\sum_{\substack{\vmh\in\Integer_+^{\th_{\nh+1}}\\ 
\vmh\equiv \vQh \mod{2}}}
q^{\frac{1}{4}\vmh\Bh\vmh-\frac{1}{2}\vAh\vmh} \frac{1}{(q)_{\mh_N}}
\prod_{\substack{j=1\\ j\neq N}}^{\th_{\nh+1}}\Mult{\mh_j+\nh_j}{\mh_j}',
\end{equation}
with $(\vmh,\vnh)$-system given by 
\begin{equation}\label{mnhat}
\vmh+\vnh=\frac{1}{2}\left(\I_{\Bh}\vmh+\vuh \right)
\end{equation}
and $\I_{\Bh}$ and $\Bh$ based on the continued fraction
expansion~\eqref{CFD}.
Notice that when $\mh_j+\nh_j\not\in\Integer$, then, 
according to \eqref{qbin_prime}, $\fh(\vuh,\vAh,\vQh)$ is zero.

\subsubsection{The case $(N/2+1)P<P'<(N+1)P$} 

The fermionic representation of $\hat{\chi}_{s,b+Nr;\ell}^{(p',p'+Np;N)}(q)$,
with $s+b+Nr+\ell$ even, 
follows from substituting $\beta_L$ of equation \eqref{BP} into the
right-hand side of the higher-level Bailey lemma~\eqref{lemma}, 
using the explicit form of $F_{r,s}^{(p,p')}(2L+\lambda,b)$.

As a first step we substitute just $\beta_L$ leaving $F_{r,s}^{(p,p')}$
unspecified. We extend the $(N-1)$-dimensional $(\vm,\vn)$-system of 
equation~\eqref{mn} to an $N$-dimensional one by setting $m_N=2L+\lambda$
and by defining
\begin{equation}\label{mn511}
\vm+\vn = \frac{1}{2}(\I_{T^{(N)}}\,\vm+\ve_{\ell}),
\end{equation}
with $\I_{T^{(N)}}$ the incidence matrix of to the tadpole
graph with $N$ nodes, 
$(\I_{T^{(N)}})_{i,j} = \delta_{|i-j|,1}+\delta_{i,j}\delta_{i,N}$.
We also define the matrix $T^{(N)}$ 
as the corresponding Cartan-type matrix $T^{(N)}=2I-\I_{T^{(N)}}$.

With this we obtain
\begin{equation}\label{chi511}
\hat{\chi}_{s,b+Nr;\ell}^{(p',p'+Np;N)}(q)=
q^{-\frac{\ell(N-\ell)+\lambda^2}{4N}}
\sum_{\substack{\vm\in\Integer_+^N\\ \vm\equiv \vQ\mod{2}}}
q^{\frac{1}{4}\vm T^{(N)} \vm} \prod_{j=1}^{N-1} \Mult{m_j+n_j}{m_j}
\frac{F_{r,s}^{(p,p')}(m_N,b)}{(q)_{m_N}},
\end{equation}
where
\begin{equation}\label{Q511}
\vQ=(\ve_{\ell+1}+\ve_{\ell+3}+\cdots)+(r+1)(\ve_1+\ve_3+\cdots).
\end{equation}
Notice that it follows from~\eqref{mn511} that $m_j+n_j>0$ if
$\vm\in\Integer_+^N$. Hence the binomials $\Mult{m_j+n_j}{m_j}$ in
\eqref{chi511} can be replaced by $\Mult{m_j+n_j}{m_j}'$ of~\eqref{qbin_prime}.
Similar arguments hold in all three cases to follow.

Next we need to substitute the fermionic polynomials.
Since $F_{r,s}^{(p,p')}$ is a linear combination of 
the elementary functions $f$ and $\tf$, we first determine the resulting 
expressions when $f$ and $\tf$ are used instead of $F_{r,s}^{(p,p')}$.

The formulas simplify if we define a new $(N+t_{n+1})$-dimensional 
$(\vmh,\vnh)$-system, combining the $(\vm,\vn)$-system~\eqref{mn511}
with the $t_{n+1}$-dimensional $(\tilde{\vm},\tilde{\vn})$-system of 
$f(m_N,\vu,\vv)$ (obtained from the equation following \eqref{f_fermi})
\begin{equation}\label{mn_f}
\tilde{\vm}+\tilde{\vn}=
\frac{1}{2}\Big( \I_B\tilde{\vm}+\vu+\vv+m_N\ve_1\Bigr).
\end{equation}
Specifically, we define $\vmh$ as
\begin{equation}\label{m_hat}
\mh_j=\begin{cases}
m_j & \text{for $1\leq j \leq N$},\\
\tilde{m}_{j-N} & \text{for $N<j\leq t_{n+1}+N$},
\end{cases}
\end{equation}
and a corresponding vector $\vnh$ through the 
$(\vmh,\vnh)$-system~\eqref{mnhat}, where the vector $\vuh$ is given by
\begin{equation}\label{uh}
\uh_j=\begin{cases} 
\delta_{j,\ell} & \text{for $1\leq j\leq N$},\\
(\vu+\vv)_{j-N} & \text{for $N<j\leq \th_{\nh+1}$}, \end{cases}
\end{equation}
and $\I_{\Bh}$ is based on the continued fraction expansion \eqref{CFD}
with $P=p'$ and $P'=p'+Np$. Using $\Bh=2I-\I_{\Bh}$ this yields
\begin{equation} \label{B511}
\Bh=\left(\begin{array}{rr|rrr}
&&&&\\&T^{(N)}\quad&&&\\&&1&&\\\hline &-1&&&\\
&&&B&\\&&&& \end{array} \right)
\end{equation}
which is in fact also the matrix one obtains for the quadratic exponent
(up to the antisymmetric part of $\Bh$ which is not fixed by 
$\vmh\Bh\vmh$). 

{}From \eqref{B511} we see that the continued fraction expansion \eqref{CFD} 
with $P=p'$ and $P'=p'+Np$ is related to that of $M(p,p')$ --used as input-- 
by 
\begin{equation}\label{nu_rel}
\nuh_0=N \qquad \text{and} \qquad \nuh_i=\nu_{i-1} \quad (1\leq i\leq \nh)
\end{equation}
where $\nh=n+1$.
Hence one additional zone (of size $N$) is added, and the 
genuine quasi-particle (corresponding to the $1/(q)_{m_N}$ in \eqref{f_hat})
sits at the last entry of this new zeroth zone.

If we further set
\begin{align}
\label{Ah}
\Ah_j&= \begin{cases} 0 & \text{for $1\leq j\leq N$},\\
(\vAh_{\vu,\vv})_{j-N} & \text{for $N<j\leq \th_{\nh+1}$}, \end{cases}\\
\intertext{and}
\label{Qh}
\Qh_j&= \begin{cases}
Q_j & \text{for $1\leq j \leq N$},\\
(\vQh_{\vu+\vv})_{j-N} & \text{for $N<j\leq \th_{\nh+1}$}, \end{cases}
\end{align}
with $\vQ$ given by \eqref{Q511}, we find that 
\begin{equation}\label{flow_BP}
f(m_N,\vu,\vv)\xrightarrow{\text{ BF }}
q^{-\frac{\ell (N-\ell )+\lambda^2}{4N}} \fh(\vuh,\vAh,\vQh),
\end{equation}
where the arrow denotes the Bailey flow obtained by inserting 
$f(m_N,\vu,\vv)$ into \eqref{chi511} instead of
$F_{r,s}^{(p,p')}(m_N,b)$.

Analogously, one finds for $\vu=\ve_{\nu_0-j-1}-\ve_{\nu_0}+\vu'$
with $0\leq j<\nu_0$ and $(\vu')_i=0$ for $1\leq i\leq \nu_0$
\begin{multline} \label{flowt_BP}
\tf(m_N,\vu,\vv) \xrightarrow{\text{ BF }} 
q^{-\frac{\ell(N-\ell)+\lambda^2}{4N}}\\ \times
\begin{cases} 
\begin{aligned}
\fh&(\vuh,\vAh_1-\ve_N,\vQh_1)
+\fh(\vuh-\ve_{N+1}-\ve_{\th_2-1}+\ve_{\th_2},\vAh_2,\vQh_2)\\
&-\fh(\vuh-\ve_{N+1}-\ve_{\th_2-1}+\ve_{\th_2},\vAh_2-2\ve_N,\vQh_2)
\end{aligned}
& \text{for $j=0$},\\[2mm]
\begin{aligned}
q^{\frac{j}{2}}&\bigl[ \fh(\vuh+\ve_{N+1}-\ve_{\th_2-j-1}
+\ve_{\th_2-j},\vAh_2,\vQh_2)\\
& -q^{\frac{1}{2}}\fh(\vuh-\ve_{\th_2-j-1}+\ve_{\th_2-j+1},
\vAh_3-\ve_N,\vQh_3) \bigr] 
\end{aligned} 
& \text{for $1\leq j<\nu_0$},
\end{cases}
\end{multline}
where $\vuh$ as in~\eqref{uh}, $\vAh_1,\vAh_2,\vAh_3$ as in~\eqref{Ah} and 
$\vQh_1,\vQh_2,\vQh_3$ as in \eqref{Qh} with $\vu$ replaced by
$\vu,\vu-\ve_{\nu_0-j-1}+\ve_{\nu_0-j},
\vu-\ve_{\nu_0-j-1}+\ve_{\nu_0-j+1}$, respectively.

Making the replacements~\eqref{flow_BP} and~\eqref{flowt_BP} 
in~\eqref{case1}--\eqref{case2b}, \eqref{case3} and~\eqref{case4} gives the 
fermionic representation of the branching function 
$\hat{\chi}_{s,b+Nr;\ell}^{(p',p'+Np;N)}(q)$.

\subsubsection{The case $P<P'<(1+N/2)P$} 
The fermionic representation of 
$\hat{\chi}_{s,b+N(b-r);\ell}^{(p',p'+N(p'-p);N)}(q)$,
with $s+b+N(b-r)+\ell$ even,
follows from substituting $\beta_L$ of equation \eqref{dBP} into the
right-hand side of the higher-level Bailey lemma~\eqref{lemma}
again using the explicit expressions for $F_{r,s}^{(p,p')}(2L+\lambda,b;1/q)$.

We follow the same strategy as before, and as an intermediate step
we leave the fermionic polynomials yet undetermined. 
We make again the variable change $m_N=2L+\lambda$ and define the
$N$-dimensional $(\vm,\vn)$-system
$\vm+\vn = \frac{1}{2}(\I_{C^{(N)}} \,\vm+\ve_{\ell})$, where
the matrices $\I_{C^{(N)}}$ and $C^{(N)}$ are the incidence and Cartan
matrix of A$_N$.
Then we obtain
\begin{multline}\label{chi512}
\hat{\chi}_{s,b+N(b-r);\ell}^{(p',p'+N(p'-p);N)}(q)=
q^{-\frac{\ell(N-\ell)+(N+1)\lambda^2}{4N}}\\
\times\sum_{\substack{\vm\in\Integer_+^N\\ \vm\equiv \vQ \mod{2}}}
q^{\frac{1}{4}\vm C^{(N)} \vm}
\prod_{j=1}^{N-1} \Mult{m_j+n_j}{m_j}
\frac{F_{r,s}^{(p,p')}(m_N,b;1/q)}{(q)_{m_N}},
\end{multline}
where
\begin{equation}\label{Q512}
\vQ=(\ve_{\ell+1}+\ve_{\ell+3}+\cdots)+(b-r)(\ve_1+\ve_3+\cdots).
\end{equation}

Similar to the previous manipulations, we now determine the result when
$F_{r,s}^{(p,p')}$ in \eqref{chi512} in replaced by $f$ and $\tf$.

Again combining the different $(\vm,\vn)$-systems, making the same variable
change as in \eqref{m_hat}, we find a new
$(N+t_{n+1})$-dimensional $(\vmh,\vnh)$-system, given by \eqref{mnhat},
with $\I_{\Bh}$ given by the continued fraction expansion \eqref{CFD}
with $P=p'$ and $P'=p'+N(p'-p)$. 
The vector $\vuh$ is still given by \eqref{uh}.
Using $\Bh=2I-\I_{\Bh}$ this yields
\begin{equation}
\Bh=\left(\begin{array}{rr|rrr}
&&&&\\&C^{(N)}\quad&&&\\&&-1&&\\\hline &-1&&&\\ 
&&&B&\\&&&& \end{array} \right)
\end{equation}
which is also the matrix that one obtains for the quadratic exponent.
This time no antisymmetric part had to be added to the quadratic form matrix
to ensure the relation $\Bh=2I-\I_{\Bh}$.

Hence the continued fraction expansion~\eqref{CFD} is this time related to 
that of the $M(p,p')$ fermionic polynomials as follows
\begin{equation}\label{nu_reld}
\nuh_0=N+\nu_0 \qquad \text{and} \qquad \nuh_i=\nu_i \quad (1\leq i\leq \nh),
\end{equation}
with $\nh=n$. 
This means that no additional zone is added, but instead the size of the 
zeroth zone has increased by $N$ so that 
the genuine quasi-particle sits in the interior of zone zero.

Further taking 
\begin{equation}\label{Ahd}
\Ah_j= \begin{cases} 
0 & \text{for $1\leq j\leq N$},\\
(\vu+\vv-\vA_{\vu,\vv})_{j-N} & \text{for $N<j\leq \th_{\nh+1}$},
\end{cases}
\end{equation}
and $\vQh$ as in \eqref{Qh} (with $\vQ$ therein being \eqref{Q512}
instead of \eqref{Q511}) we find that
\begin{equation} \label{flowd_BP}
f(m_N,\vu,\vv;1/q)\xrightarrow{\text{ BF }}
q^{-\frac{\ell (N-\ell )+(N+1)\lambda^2}{4N}} \fh(\vuh,\vAh,\vQh).
\end{equation}

The analogous transformation for $\tf$ is given by
\begin{multline}\label{flowdt_BP}
\tf(m_N,\vu,\vv;1/q)
\xrightarrow{\text{ BF }} 
q^{-\frac{\ell(N-\ell)+(N+1)\lambda^2}{4N}}\\ \times
\begin{cases} 
\begin{aligned}
\fh&(\vuh,\vAh_1+\ve_N,\vQh_1)
+\fh(\vuh-\ve_{\th_1-1}+\ve_{\th_1}-\ve_{N+1},\vAh_2,\vQh_2)\\
&-\fh(\vuh-\ve_{\th_1-1}+\ve_{\th_1}-\ve_{N+1},\vAh_2+2\ve_N,\vQh_2)
\end{aligned}
& \text{for $j=0$},\\[2mm]
\begin{aligned}
q^{-\frac{j}{2}}&\bigl[ \fh(\vuh-\ve_{\th_1-j-1}+\ve_{\th_1-j}+\ve_{N+1},
\vAh_2,\vQh_2)\\
& -q^{-\frac{1}{2}}\fh(\vuh-\ve_{\th_1-j-1}+\ve_{\th_1-j+1},\vAh_3+\ve_N,
\vQh_3) \bigr]
\end{aligned} 
& \text{for $1\leq j<\nu_0$},
\end{cases}
\end{multline}
where $\vuh$ as in~\eqref{uh}, $\vAh_1,\vAh_2,\vAh_3$ as in~\eqref{Ahd} and 
$\vQh_1,\vQh_2,\vQh_3$ as in \eqref{Qh} with $\vu$ replaced by
$\vu,\vu-\ve_{\nu_0-j-1}+\ve_{\nu_0-j},
\vu-\ve_{\nu_0-j-1}+\ve_{\nu_0-j+1}$, respectively.

Making the replacements~\eqref{flowd_BP} and~\eqref{flowdt_BP} 
in~\eqref{case1}-\eqref{case2b}, \eqref{case3} and~\eqref{case4} gives the 
fer\-mionic form of the branching function 
$\hat{\chi}_{s,b+N(b-r);\ell}^{(p',p'+N(p'-p);N)}(q)$.

\subsection{Fermionic Branching Functions for $(P,P')_N$ 
with $P'>(N+1)P$}\label{sec_big} 

To proceed in a fashion similar to section~\ref{sec_small}
we now introduce the elementary fermionic function
\begin{equation}\label{def_gh}
\gh(\vuh,\vAh,\vQh)=\sum_{\substack{\vmh\in\Integer_+^{\th_{\nh+1}}\\
\vmh\equiv \vQh \mod{2}}}
\frac{q^{\frac{1}{4}\vmh \Bh\vmh-\frac{1}{2}\vAh\vmh}}
{(q)_{\nh_{\th_1+1}}\cdots(q)_{\nh_{\th_2}}(q)_{\mh_{\th_2+1}}}
\prod_{\substack{j=1\\j\neq \th_1+1,\ldots,\th_2+1}}^{\th_{\nh+1}}
\Mult{\mh_j+\nh_j}{\mh_j}',
\end{equation}
with $(\vmh,\vnh)$-system given by \eqref{mnhat}, where $\I_{\Bh}$ 
and $\Bh=2I-\I_{\Bh}$ follow from the continued fraction 
expansion \eqref{CFD} as before.
Again, $\gh(\vuh,\vAh,\vQh)$ is zero if $\mh_j+\nh_j\not\in\Integer$.
In contrast to \eqref{f_hat}, however, some of the genuine quasi-particles
corresponding to the factors $1/(q)_a$ are labelled by $a=\nh_j$ instead 
of $a=\mh_j$.
Note that $\nh_j$, determined from \eqref{mnhat}, can take negative values,
and we adopt the convention here that $1/(q)_{-n}=0$ for $n>0$.

\subsubsection{The case $(Nk+N/2+1)P<P'<(Nk+N+1)P$ with $k\geq 1$}
The fermionic representation of 
$\hat{\chi}_{s,b+N(kb+r);\ell}^{(p',p'+N(p+kp');N)}(q)$,
with $s+b+N(kb+r)+\ell$ even,
follows from substituting $\beta_L^{(k)}$ of equation \eqref{BPchain} into the
right-hand side of the higher-level Bailey lemma~\eqref{lemma},
using the explicit form of $F_{r,s}^{(p,p')}(2r_k+\lambda,b)$.

As before, we first give the results when $F_{r,s}^{(p,p')}$ is still 
unspecified. We extend the $(N-1)$-dimensional $(\vm,\vn)$-system of the 
Bailey lemma by defining
\begin{equation}\label{mn_ext}
\begin{split}
n_j&=\begin{cases}
L-r_1 & \text{for $j=N$},\\
r_{j-N}-r_{j-N+1} & \text{for $N<j<N+k$},
\end{cases}\\
m_{N+k}&=2r_k+\lambda
\end{split}
\end{equation}
and setting the $(N+k)$-dimensional $(\vm,\vn)$-system
\begin{equation}\label{mn521}
\vm+\vn=\frac{1}{2}\bigl(\I_{B^{(N+k)}}\vm+\ve_{\ell}\bigr).
\end{equation}
The matrix $\I_{B^{(N+k)}}$ is defined by \eqref{I_B} with $\nu_0=N-1$,
$\nu_1=k$, $\nu_2=1$ and $\nu_3=0$ and $B^{(N+k)}=2I-\I_{B^{(N+k)}}$.
The reason for labelling some of the variables in \eqref{mn_ext} by
$n_j$ is that these do not have any parity restrictions. This way we
also ensure that the $(\vm,\vn)$-system \eqref{mn521} is based on a
fractional-level incidence matrix of the form \eqref{I_B}.

With these variable changes we obtain
\begin{multline}\label{chi521}
\hat{\chi}_{s,b+N(kb+r);\ell}^{(p',p'+N(p+kp');N)}(q)
=q^{-\frac{\ell(N-\ell)+(Nk+1)\lambda^2}{4N}}\\
\times\sum_{\substack{\vm\in\Integer_+^{N+k}\\ \vm\equiv \vQ \mod{2}}}
q^{\frac{1}{4}\vm B^{(N+k)}\vm} 
\prod_{j=1}^{N-1} \Mult{m_j+n_j}{m_j}
\frac{F_{r,s}^{(p,p')}(m_{N+k},b)}{(q)_{n_N}\cdots (q)_{n_{N+k-1}}
(q)_{m_{N+k}}},
\end{multline}
where 
\begin{equation}
Q_j=\begin{cases}
(\delta_{\ell+1,j}+\delta_{\ell+3,j}+\cdots)
+(kb+r+1)(\delta_{j,1}+\delta_{j,3}+\cdots) 
& \text{for $1\leq j<N$},\\
0 & \text{for $N\leq j<N+k$},\\
\lambda & \text{for $j=N+k$}.
\end{cases}
\end{equation}

Next we again substitute $f$ and $\tilde{f}$ for $F_{r,s}^{(p,p')}$
into \eqref{chi521}. The resulting formulas can be simplified by combining
the $(\vm,\vn)$-system~\eqref{mn521} with the $t_{n+1}$-dimensional
$(\tilde{\vm},\tilde{\vn})$-system of $f(m_{N+k},\vu,\vv)$
\begin{equation}
\tilde{\vm}+\tilde{\vn}=\frac{1}{2}\left( \I_B\tilde{\vm}+\vu+\vv
+m_{N+k}\ve_1\right)
\end{equation}
as
\begin{equation} \label{mn_new_hat}
\mh_j=\begin{cases}
m_j & \text{for $1\leq j\leq N+k$},\\
\tilde{m}_j & \text{for $N+k<j\leq t_{n+1}+N+k$},
\end{cases}
\end{equation}
and a corresponding vector $\vnh$ through the 
$(\vmh,\vnh)$-system~\eqref{mnhat}, where
\begin{equation}\label{uh_1}
\uh_j=\begin{cases}
\delta_{j,\ell} & \text{for $1\leq j<N$},\\
0 & \text{for $N\leq j\leq N+k$},\\
(\vu+\vv)_{j-N-k} &\text{for $ N+k<j\leq N+k+t_{n+1}$}.
\end{cases}
\end{equation}
and $\I_{\Bh}$ is based on the continued fraction expansion \eqref{CFD} with 
$P=p'$ and $P'=p'+N(p+kp')$.
Using $\Bh=2I-\I_{\Bh}$ this yields
\begin{equation} \label{B521}
\Bh=\left(\begin{array}{rr|rr|r|rrr}
&&&&&&&\\&T^{(N-1)}\,&&&&&&\\&&1&&&&&\\\hline &-1&&&&&&\\
&&&T^{(k)}\quad &&&&\\&&&&1&&&\\\hline &&&-1&1&1&&\\\hline &&&&-1&&&\\
&&&&&&B&\\&&&&&&& \end{array}\right)
\end{equation}
which is indeed also the matrix one obtains from the quadratic exponent by
noticing that the top-left $(N+k)\times (N+k)$ entries
of $\Bh$ correspond to $B^{(N+k)}$. 

{}From \eqref{B521} we see that $\Bh$ is based on the 
continued fraction expansion \eqref{CFD} with $P=p'$ and $P'=p'+N(p+kp')$
related to the continued fraction expansion of $M(p,p')$ by
\begin{equation} \label{nu_rel1}
\nuh_0=N-1,~~\nuh_1=k,~~\nuh_2=1 \quad \text{and} \quad 
\nuh_i=\nu_{i-3}~~\text{for $3\leq i\leq \nh$}
\end{equation}
with $\nh=n+3$, which means that three additional zones have been added. 

Upon further setting
\begin{align}\label{Ah_1}
\Ah_j&=\begin{cases}
0 & \text{for $1\leq j\leq \th_2+1$},\\
(\vA_{\vu,\vv})_{j-\th_2-1} &\text{for $ \th_2+1<j\leq \th_{\nh+1}$},\\
\end{cases}\\
\intertext{and}
\label{Qh_1}
\Qh_j&=\begin{cases}
Q_j & \text{for $1\leq j\leq \th_2+1$},\\
(\vQ_{\vu+\vv})_{j-\th_2-1} &\text{for $\th_2+1<j\leq \th_{\nh+1}$},
\end{cases}
\end{align}
we obtain
\begin{equation}\label{flow_BPchain}
f(m_{N+k},\vu,\vv)\xrightarrow{\text{ BF }}
q^{-\frac{\ell (N-\ell)+(Nk+1)\lambda^2}{4N}} 
\gh(\vuh,\vAh,\vQh)
\end{equation}
and similarly
\begin{multline}\label{flowt_BPchain}
\tilde{f}(2r_k+\lambda,\vu,\vv) \xrightarrow{\text{BF}}
q^{-\frac{\ell(N-\ell)+(Nk+1)\lambda^2}{4N}}\\
\times \begin{cases}
\begin{split}
\gh&(\vuh,\vAh_1-\ve_{\th_2+1},\vQh_1)\\
&+\gh(\vuh-\ve_{\th_2+2}-\ve_{\th_4-1}+\ve_{\th_4},\vAh_2,\vQh_2)\\
&-\gh(\vuh-\ve_{\th_2+2}-\ve_{\th_4-1}+\ve_{\th_4},\vAh_2-2\ve_{\th_2+1},
\vQh_2)
\end{split} & \text{for $j=0$},\\
\begin{split}
q^{\frac{j}{2}}\bigl[ &\gh(\vuh+\ve_{\th_2+2}-\ve_{\th_4-j-1}+\ve_{\th_4-j},
\vAh_2,\vQh_2)\\
&-q^{\frac{1}{2}}\gh(\vuh-\ve_{\th_4-j-1}+\ve_{\th_4-j+1},
\vAh_3-\ve_{\th_2+1},\vQh_3) \bigr]
\end{split} & \text{for $1\leq j<\nu_0$},
\end{cases}
\end{multline}
where $\vuh$ as in~\eqref{uh_1}, $\vAh_1,\vAh_2,\vAh_3$ as in~\eqref{Ah_1} and 
$\vQh_1,\vQh_2,\vQh_3$ as in \eqref{Qh_1} with $\vu$ replaced by
$\vu,\vu-\ve_{\nu_0-j-1}+\ve_{\nu_0-j},
\vu-\ve_{\nu_0-j-1}+\ve_{\nu_0-j+1}$, respectively.

Making the replacements~\eqref{flow_BPchain} and~\eqref{flowt_BPchain} 
in~\eqref{case1}--\eqref{case2b}, \eqref{case3} and~\eqref{case4} gives the 
fermionic expression for the branching function 
$\hat{\chi}_{s,b+N(kb+r);\ell}^{(p',p'+N(p+kp');N)}(q)$.

\subsubsection{The case $(Nk+1)P<P'<(Nk+N/2+1)P$ with $k\geq 1$} 
The fermionic representation of
$\hat{\chi}_{s,b+N(kb+b-r);\ell}^{(p',p'+N(kp'+p'-p);N)}(q)$
with $s+b+N(kb+b-r)+\ell$ even,
follows from substituting $\beta_L^{(k)}$ of equation \eqref{dBPchain} into the
right-hand side of the higher-level Bailey lemma~\eqref{lemma}
using the explicit form of $F_{r,s}^{(p,p')}(2r_k+\lambda,b;1/q)$.
We first obtain
\begin{multline}\label{chi522}
\hat{\chi}_{s,b+N(kb+b-r);\ell}^{(p',p'+N(kp'+p'-p);N)}(q)
=q^{-\frac{\ell(N-\ell)+(Nk+N+1)\lambda^2}{4N}}\\
\times\sum_{\substack{\vm\in\Integer_+^{N+k}\\ \vm\equiv \vQ \mod{2}}}
q^{\frac{1}{4}\vm B^{(N+k)}\vm} 
\prod_{j=1}^{N-1} \Mult{m_j+n_j}{m_j}
\frac{F_{r,s}^{(p,p')}(m_{N+k},b;1/q)}{(q)_{n_N}\cdots (q)_{n_{N+k-1}}
(q)_{m_{N+k}}},
\end{multline}
where 
\begin{equation} \label{Q_new}
Q_j=\begin{cases}
(\delta_{\ell+1,j}+\delta_{\ell+3,j}+\cdots)
+(kb+b-r+1)(\delta_{j,1}+\delta_{j,3}+\cdots) 
& \text{for $1\leq j<N$},\\
0 & \text{for $N\leq j<N+k$},\\
\lambda & \text{for $j=N+k$}.
\end{cases}
\end{equation}
The $(N+k)$-dimensional $(\vm,\vn)$-system is that of equation~\eqref{mn521},
but this time with $\I_{B^{(N+k)}}$ (and thus $B^{(N+k)}$) based on the
continued fraction expansion with $\nu_0=N-1$,
$\nu_1=k$ and $\nu_2=1$.
We note that to get to this result the same variable change as in
\eqref{mn_ext} has been carried out.

We now again define variables as in \eqref{mn_new_hat} and the 
corresponding $(\vmh,\vnh)$-system \eqref{mnhat}
with $\vuh$ defined as in \eqref{uh_1} and $\I_{\Bh}$
this time based on the continued fraction expansion \eqref{CFD}
with $P=p'$ and $P'=p'+N(kp'+p'-p)$. This yields the matrix 
\begin{equation}
\Bh=\left(\begin{array}{rr|rr|rlr}
&&&&&&\\&T^{(N-1)}\,&&&&&\\&&1&&&&\\\hline &-1&&&&&\\
&&&T^{(k)}\quad &&&\\&&&&1&&\\\hline &&&-1&2&-1&\\ &&&&-1&&\\
&&&&&\quad B&\\&&&&&& \end{array}\right)
\end{equation}
which again agrees with the matrix one obtains from the quadratic exponent
(up to arbitrary antisymmetric pieces).

The old and new continued fraction expansions are related by
\begin{equation}\label{nu_reld1}
\nuh_0=N-1,~~\nuh_1=k,~~\nuh_2=\nu_0+1 \quad \text{and} \quad 
\nuh_i=\nu_{i-2}~~\text{for $3\leq i\leq \nh$}
\end{equation}
with $\nh=n+2$. This corresponds to the addition of 2 extra zones.

We further define $\vQh$ as in \eqref{Qh_1} with $Q_j$ as in \eqref{Q_new} and 
\begin{equation}\label{Ahd_1}
\Ah_j=\begin{cases}
0 & \text{for $1\leq j\leq \th_2+1$},\\
(\vu+\vv-\vA_{\vu,\vv})_{j-\th_2-1} & 
\text{for $\th_2+1<j\leq \th_{\nh+1}$}.
\end{cases}
\end{equation}
Then 
\begin{equation}
\label{flowd_BPchain}
f(2r_k+\lambda,\vu_b,\vu_s;1/q)\xrightarrow{\text{ BF }}
q^{-\frac{\ell(N-\ell)+(Nk+N+1)\lambda^2}{4N}} \gh(\vuh,\vAh,\vQh)
\end{equation}
and 
\begin{multline}\label{flowdt_BPchain}
\tilde{f}(2r_k+\lambda,\vu,\vv;1/q)
\xrightarrow{\text{ BF }}
q^{-\frac{\ell(N-\ell)+(Nk+N+1)\lambda^2}{4N}}\\
\times \begin{cases}
\begin{split}
\gh&(\vuh,\vAh_1+\ve_{\th_2+1},\vQh_1)
+\gh(\vuh-\ve_{\th_2+2}-\ve_{\th_3-1}+\ve_{\th_3},\vAh_2,\vQh_2)\\
&-\gh(\vuh-\ve_{\th_2+2}-\ve_{\th_3-1}+\ve_{\th_3},\vAh_2+2\ve_{\th_2+1},
\vQh_2)
\end{split}
& \text{for $j=0$},\\
\begin{split}
q^{-\frac{j}{2}}\bigl[ &\gh(\vuh+\ve_{\th_2+2}-\ve_{\th_3-j-1}+\ve_{\th_3-j},
\vAh_2,\vQh_2)\\
&-q^{-\frac{1}{2}}\gh(\vuh-\ve_{\th_3-j-1}+\ve_{\th_3-j+1},
\vAh_3+\ve_{\th_2+1},\vQh_3)
\end{split}
& \text{for $1\leq j<\nu_0$},
\end{cases}
\end{multline}
where $\vuh$ as in~\eqref{uh_1}, $\vAh_1,\vAh_2,\vAh_3$ as in~\eqref{Ahd_1} 
and $\vQh_1,\vQh_2,\vQh_3$ as in \eqref{Qh_1} with $\vu$ replaced by
$\vu,\vu-\ve_{\nu_0-j-1}+\ve_{\nu_0-j},
\vu-\ve_{\nu_0-j-1}+\ve_{\nu_0-j+1}$, respectively.

Making the replacements~\eqref{flowd_BPchain} and~\eqref{flowdt_BPchain} 
in~\eqref{case1}--\eqref{case2b}, \eqref{case3} and~\eqref{case4} gives the 
fermionic form of $\hat{\chi}_{s,b+N(kb+b-r);\ell}^{(p',p'+N(kp'+p'-p);N)}(q)$.

\vspace{2mm}
To conclude section \ref{sec_big}, let us remark that for $N=1$ the continued 
fraction expansion used in~\eqref{CFD} is not the same as the one used in 
refs.~\cite{BM96,BMS96} for the case $P'>2P$. Here we considered the 
continued fraction expansion of $P'/(P'-P)$ whereas in~\cite{BM96,BMS96} that 
of $P'/P$ was used. The difference between these two cases is that in the 
first one a zero zone of length zero ({\it i.e.} $\nuh_0=0$) is obtained
whereas the second case starts with $\nuh_0'=\nuh_1$. This of course changes 
the Takahashi decomposition, but the final results are the same (in particular
the fractional-level incidence matrices for the $(\vmh,\vnh)$-system are the 
same).

\section{Discussion} 
\label{sec_discussion}

In the introduction we briefly discussed the  relation of the 
massless RG flow of \cite{ZAMA}-\cite {ZAMB}
to Bailey flow. Specifically we noted that the RG flow of ~\cite{ALZAM}
flows with a continuous 
parameter from $M(p,p+1)$ to $M(p-1,p)$ whereas, in contrast,  Bailey
flow is a discrete
process which adds degrees of freedom and, in this special case, goes from 
$M(p-1,p)$ to $M(p,p+1).$ 
Moreover the parameter $L$ which appears in the polynomial identities
for $M(p,p')$ is, as emphasized by Melzer~\cite{mel}, an ultraviolet
cutoff, which certainly is reminiscent of the RG flow phenomena.
However the precise (if any) connection between the two concepts is
not yet understood and we will conclude this paper by discussing a few
of the possibilities. 

The special case of $M(p,p+1)$ might suggest that the Bailey construction 
could be generalized to include a free parameter so that it would be an exact
inverse to RG flow. This is particularly the case with the flow
between $M(3,4)$ and $M(4,5).$ 
Here, using the equivalence of $M(4,5)$ with
the $N=1$ supersymmetric model $SM(3,5)$, the Bailey 
construction of~\cite{BMS96a}, which utilizes the original lemma of
Bailey~\eqref{B_lemma} with $\rho_1\to\infty$ provides a one parameter
flow (in the variable $\rho_2$) from $SM(3,5)=M(4,5)$ to $M(3,4)$.
The model $SM(3,5)=M(4,5)$ corresponds to $\rho_2=q^{1/2}$ or $q$ and
$M(3,4)$ to $\rho_2\to \infty$.
The $q\rightarrow 1$ limit of the vacuum character has been studied 
in~\cite{Chim} and this provides a function which satisfies the
property of the $c$ function of~\cite{ZAMB} in interpolating between the two
models.

Furthermore, in~\cite{BMS96a} the full two parameters in Bailey's 
lemma~\eqref{B_lemma} are exploited to give flows from the
$N=2$ supersymmetric models with central charge $c=3(1-2/p)$ to
$SM(p,p+2)$ to $M(p,p+1)$. The corresponding $q\rightarrow 1$ limit of
the vacuum character is computed in~\cite{Chim} and this also
interpolates between the central charges of the models. In
this case no RG flow (or TBA) analysis with two continuous parameters
has been carried out.

On the other hand it might be expected that the TBA equations 
of~\cite{ALZAM} which give the one
parameter RG flow $M(p,p+1)\rightarrow M(p-1,p)$ should have an
extension to a one parameter flow of characters. But it seems to us
that if this is true it may require further identities that are not yet
extant. The reason for this is that on the fermionic side
the Bailey constructions add particles on one side of the genuine quasi
particles (those which contribute $1/(q)_n$) and 
the tail of ghost (or pseudo) particles which
correspond to the $q$-binomials lies on the other side of the genuine
quasi particles. This leads to asymmetric fermionic representations of
the characters  whereas the
TBA equations of~\cite{ALZAM} treat the two ends of the equations in a
symmetric fashion. This problem needs to be explored.  

We also note that the Bailey flows of this paper of $M(p,p')$ into
the cosets~\eqref{coset} is not the inverse of the flow  of~\cite{CSS,Zam91} 
which is between different cosets of the form~\eqref{coset} and does not 
in general involve $M(p,p')$. 

It thus seems that Bailey flow and RG flow give somewhat
different relations between models of CFT. However what is lacking at
present in the method of Bailey flow is an abstract understanding of
why Bailey's construction is related to CFT at all. At
present we are only able to make the identification of the Bailey flow
characters with the CFT characters by comparing the results of two
separate computations. Since all examples of Bailey flow have been
identified with CFT models this cannot be an accidental coincidence
and it is most desirable to prove that there is a
connection between the Bailey flow and CFT which allows us to identify
the CFT model without the need of doing a separate Feigin-Fuchs computation
of the CFT characters. Such a theorem would allow the Bailey
flow to be a complete constructive procedure which could
serve as an alternative route to the construction of CFT.

\section*{Acknowledgements}
This work was partially supported by NSF grants DMR9404747 and 
DMS9501101 and the Australian Research Council.

\appendix
\section*{-- Appendix --}
\section*{A$^{(1)}_1$ bosonic branching functions}  
\setcounter{section}{1}

The bosonic form of the branching functions for the coset model $(P,P')_N$ 
with $N\geq 2$
been calculated in~\cite{D87}--\cite{R} for integer levels $N$ and $N'$ 
(unitary models). The cosets~\eqref{coset} with fractional $N'$ have been
considered in \cite{Ahn91}.
Ahn {\em et al.}~\cite{ACT} determined the branching functions for
fractional levels and specializing their results to integer $N$,
we find (for all $N \geq 1$)
\begin{equation}\label{bfN1}
\chi_{r,s;\ell}^{(P,P';N)}(q)=q^{-\frac{c}{24}+\Delta_{r,s}
+\frac{\ell(N-\ell)}{2N(N+2)}}\hat{\chi}_{r,s;\ell}^{(P,P';N)}(q)
\end{equation}
with the normalized branching function
\begin{multline}\label{bfN}
\hat{\chi}_{r,s;\ell}^{(P,P';N)}(q)=q^{-\frac{\ell(N-\ell)}{2N(N+2)}}\\
\times
\sum_{0\leq m \leq N/2} c_{2m}^{\ell}(q) \biggl(
\sum_{\substack{j \in \Integer\\ m_{r-s}(j)\equiv \pm m \mod{N}}} 
\hspace{-6mm}  q^{\frac{j}{N}(jPP'+P'r-Ps)}
- \hspace{-4mm} \sum_{\substack{j\in \Integer \\ 
m_{r+s}(j)\equiv \pm m \mod{N} }} 
\hspace{-6mm} q^{\frac{1}{N}(jP'+s)(jP+r)} \biggr).
\end{multline}
Here $1\leq r \leq P-1,~1\leq s\leq P'-1$, $0\leq \ell \leq N$, and 
$r-s$ and $\ell$ are either both even or both odd.
The first sum in (\ref{bfN}) runs over integer $m$ when $\ell$ is even 
and half an odd integer $m$ when $\ell$ is odd, generalizing the 
Neveu--Schwarz 
and Ramond sector of the supersymmetric case corresponding to $N=2$. 
The restriction on $m_a(j):=(a/2+P'j)$ in the sum over $j$ indicates that we 
only sum over those values of $j$ for which $m_a(j)\equiv \pm m \mod N$.

The central charge and conformal dimensions in (\ref{bfN1}) are given by
\begin{align}
c&=\frac{3N}{N+2}\left(1-\frac{2(N+2)}{N^2}\frac{(P'-P)^2}{P'P}\right)
=\frac{3NN'(N'+N+4)}{(N+N'+2)(N+2)(N'+2)} \\
&= 1- \frac{6N}{(N'+2)(N'+N+2)}+\frac{2(N-1)}{N+2}
\label{ccN}
\end{align}
and
\begin{equation}
\label{cdN}
\Delta_{r,s}=\frac{(P'r-Ps)^2-(P'-P)^2}{4NP'P}
=\frac{[ (N+N'+2)r-(N'+2)s]^2-N^2}{4N(N+N'+2)(N'+2)}.
\end{equation}
Expression (\ref{ccN}) for the central charge reflects that 
in the Feigin and Fuchs construction one deals with a Z$_N$-parafermion field
with central charge equal to the third term in (\ref{ccN}) and a bosonic 
field with a background charge with central charge given by the 
first two terms in (\ref{ccN}).

The function $c_m^{\ell}$ in (\ref{bfN}) is the level-$N$
A$^{(1)}_1$ string function~\cite{KP84,JM84}.
We need in section~\ref{sec_Bflow} the form given by Lepowsky and 
Primc~\cite{LP85} obtained by using $Z$ algebras
\begin{equation}
\label{sf}
c_m^{\ell}(q)=\frac{q^{\frac{\ell(N-\ell)}{2N(N+2)}}}{(q)_{\infty}}
\sum_{\substack{\vn \in \Integer_+^{N-1} \\ 
\frac{m+\ell}{2N}+(C^{-1}\vn)_1 \in \Integer}}
\frac{q^{\vn C^{-1}(\vn-\ve_{\ell})}}
{(q)_{n_1}\ldots (q)_{n_{N-1}}},
\end{equation}
and also the symmetry properties
\begin{equation}
\label{sfsym}
c_m^{\ell}=c_{-m}^{\ell}=c_{m+2N}^{\ell}=c_{N-m}^{N-\ell}.
\end{equation}


\begin{thebibliography}{50}
\bibitem{ZAMA}
A.~B.~Zamolodchikov, {\it Sov.\ J. Nucl.\ Phys.} {\bf 46}, 1090 (1987).
\bibitem{LC} 
A.~Ludwig and J.~Cardy, {\it Nucl.\ Phys.} {\bf B287} [FS19], 687 (1987).
\bibitem{ALZAM}
Al.~B.~Zamolodchikov, {\it Nucl.\ Phys.} {\bf B358}, 524 (1991).
\bibitem{AHN}
C.~Ahn, {\it Phys.\ Letts.} {\bf B294}, 204 (1992).
\bibitem{CSS}
C.~Crnkovic, G.~M.~Sotkov and M.~Stanishkov, {\it Phys.\ Letts.} 
{\bf B226}, 297 (1989).
\bibitem{Zam91}
Al.~B.~Zamolodchikov, {\it Nucl.\ Phys.} {\bf B366}, 122 (1991).
\bibitem{ZAMB}
A.~B.~Zamolodchikov, {\it JETP Letts.} {\bf 43}, 730 (1986).
\bibitem{FQ94}
O.~Foda and Y.-H.~Quano,
``Virasoro character identities from the
Andrews--Bailey construction'',
to appear in {\it Int.\ J. Mod.\ Phys.\ A}, hep-th/9408086.
\bibitem{Ole2}
S.~O.~Warnaar,
{\it J. Stat.\ Phys.} {\bf 84}, 49 (1996).
\bibitem{BMS96a}  A.~Berkovich, B.~M.~McCoy and A.~Schilling,
{\it Physica A} {\bf 228}, 33 (1996).
\bibitem{Bailey}
W.~N.~Bailey,
{\it Proc.\ London Math.\ Soc.}(2) {\bf 50}, 1 (1949).
\bibitem{Slater51}
L.~J.~Slater,
{\it Proc.\ Lond.\ Math.\ Soc.} (2) {\bf 52}, 460 (1951).
\bibitem{Slater52}
L.~J.~Slater,
{\it Proc.\ Lond.\ Math.\ Soc.} (2) {\bf 54}, 147 (1951-52).
\bibitem{Rogers94}
L.~J.~Rogers,
{\it Proc.\ Lond.\ Math.\ Soc.} {\bf 25}, 318 (1894).
\bibitem{Rogers17}
L.~J.~Rogers,
{\it Proc.\ Lond.\ Math.\ Soc.} (2) {\bf 16}, 315 (1917).
\bibitem{Andrews75}
G.~E.~Andrews,
in: ``The Theory and Application of Special Functions'', R.~Askey, ed.
(Academic Press, New York) pp. 191 (1975).
\bibitem{Andrews84}
G.~E.~Andrews,
{\it Pacific J. Math.} {\bf 114}, 267 (1984).
\bibitem{Paule}
P.~Paule,
{\it J. Math.\ Anal.\ Appl.} {\bf 107}, 255 (1985).
\bibitem{Milne92}
S.~C.~Milne and G.~M.~Lilly,
{\it Bull.\ Amer.\ Math.\ Soc.} (N.S.) {\bf 26}, 258 (1992).
\bibitem{Lilly93}
G.~M.~Lilly and S.~C.~Milne,
{\it Constr.\ Approx.} {\bf 9}, 473 (1993).
\bibitem{Milne95}
S.~C.~Milne and G.~M.~Lilly,
{\it Discrete Math.} {\bf 139}, 319 (1995).
\bibitem{SW96hl1} 
A.~Schilling and S.~O.~Warnaar,
{\it Int.\ J. Mod.\ Phys.} {\bf B11}, 189 (1997).
\bibitem{SW96hl2} 
A.~Schilling and S.~O.~Warnaar,
``A higher-level Bailey lemma: Proof and application'',
to appear in {\it the Ramanujan Journal}, q-alg/9607014.
\bibitem{Ram}
L.~J.~Rogers and S.~Ramanujan, 
{\it Proc.\ Camb.\ Phil.\ Soc.} {\bf 19}, 211 (1919).
\bibitem{Dobrev}
V.~K.~Dobrev,
{\it Supplemento ai Rendiconti del Circolo Matematico di Palermo},
Serie II, numero 14, 25 (1987).
\bibitem{CIZ87}
A.~Capelli, C.~Itzykson and J.~B.~Zuber,
{\it Commun.\ Math.\ Phys.} {\bf 113}, 1 (1987).
\bibitem{FF83} 
B.~Feigin and D.~B.~Fuchs,
{\it Funct.\ Anal.\ Appl.} {\bf 17}, 241 (1983).
\bibitem{KKMM1}
R.~Kedem, T.~R.~Klassen, B.~M.~McCoy and E.~Melzer,
{\it Phys.\ Lett.} {\bf B304}, 263 (1993).
\bibitem{KKMM2}
R.~Kedem, T.~R.~Klassen, B.~M.~McCoy and E.~Melzer,
{\it Phys.\ Lett.} {\bf B307}, 68 (1993).
\bibitem{Kent}
A.~Kent,
Ph.D. Thesis, Cambridge University (1986).
\bibitem{RC83}
A.~Rocha--Caridi,
``Vacuum vector representations of the Virasoro algebra'' in
{\it Vertex Operators in Mathematics and Physics}, Springer-Verlag (1984). 
\bibitem{D87}
E.~Date, M.~Jimbo, A.~Kuniba, T.~Miwa and M.~Okado,
{\it Nucl.\ Phys.} {\bf B290}, 231 (1987);\\
E.~Date, M.~Jimbo, T.~Miwa and M.~Okado,
{\it Phys.\ Rev.} {\bf B35}, 2105 (1987).
\bibitem{D88}
E.~Date, M.~Jimbo, A.~Kuniba, T.~Miwa and M.~Okado,
{\it Adv.\ Stud.\ in Pure Math.} {\bf 16}, 17 (1988).
\bibitem{KMQ} 
D.~Kastor, E.~Martinec and Z.~Qiu,
{\it Phys.\ Lett.} {\bf B200}, 434 (1988).
\bibitem{BNY} 
J.~Bagger, D.~Nemeschansky and S.~Yankielowicz,
{\it Phys.\ Rev.\ Lett.} {\bf 60}, 389 (1988).
\bibitem{R} 
F.~Ravanini,
{\it Mod.\ Phys.\ Lett.} {\bf A3}, 271 and 397 (1988).
\bibitem{KW88}
V.~G.~Kac and M.~Wakimoto,
{\it Proc.\ Natl.\ Acad.\ Sci.\ USA} {\bf 85}, 4956 (1988).
\bibitem{Ahn91}
C.~Ahn,
{\it Nucl.\ Phys.} {\bf B354}, 57 (1991).
\bibitem{ACT} 
C.~Ahn, S.-W.~Chung and S.-H.~Tye,
{\it Nucl. Phys.} {\bf B365}, 191 (1991).
\bibitem{BEHHH}
R.~Blumenhagen, W.~Eholzer, A.~Honecker, K.~Hornfeck and R.~Huebel,
{\it Int.\ J.\ Mod.\ Phys.} {\bf A10}, 2367 (1995).
\bibitem{BM96} 
A.~Berkovich and B.~M.~McCoy,
{\it Lett.\ Math.\ Phys.} {\bf 37}, 49 (1996).
\bibitem{BMS96}
A.~Berkovich, B.~M.~McCoy and A.~Schilling,
``Rogers--Schur--Ramanujan type identities for the $M(p,p')$ minimal models
of conformal field theory'', submitted to {\it Comm.\ Math.\ Phys.},
q-alg/9607020.
\bibitem{S96a}
A.~Schilling,
{\it Nucl.\ Phys.} {\bf B459}, 393 (1996).
\bibitem{S96b}
A.~Schilling,
{\it Nucl.\ Phys.} {\bf B467}, 247 (1996).
\bibitem{W}
S.~O.~Warnaar,
``The Andrews--Gordon identities and $q$--multinomial coefficients'',
to appear in {\it Comm.\ Math.\ Phys.}, q-alg/9601012.
\bibitem{SW97}
A.~Schilling and S.~O.~Warnaar,
``Supernomial coefficients, polynomial identities and $q$--series'', 
submitted to {\it the Ramanujan Journal}, q-alg/9701007.
\bibitem{ABF}
G.~E.~Andrews, R.~J.~Baxter and P.~J.~Forrester,
{\it J. Stat.\ Phys.} {\bf 35}, 193 (1984).
\bibitem{FB} 
P.~J.~Forrester and R.~J.~Baxter,
{\it J. Stat.\ Phys.} {\bf 38}, 435 (1985).
\bibitem{Bressoud}
D.~M.~Bressoud,
``The Bailey lattice: An introduction'', pp. 57-67 in {\it Ramanujan
Revisited.} G.~E.~Andrews {\it et al.} eds., Academic Press (1988).
\bibitem{NY}
A.~Nakayashiki and Y.~Yamada,
{\it Int.\ J. Mod.\ Phys. A} {\bf 11}, 395 (1996).
\bibitem{FQ95}
O.~Foda and Y.-H.~Quano,
{\it Int.\ J. Mod.\ Phys. A} {\bf 10}, 2291 (1995).
\bibitem{Kir}
A.~N.~Kirillov,
{\it Prog.\ Theor.\ Phys.\ Suppl.} {\bf 118}, 61 (1995).
\bibitem{B94}
A.~Berkovich,
{\it Nucl.\ Phys.} {\bf B431}, 315 (1994).
\bibitem{FW}
O.~Foda and S.~O.~Warnaar,
{\it Lett.\ Math.\ Phys.} {\bf 36}, 145 (1996).
\bibitem{Ole1}
S.~O.~Warnaar,
{\it J. Stat.\ Phys.} {\bf 82}, 657 (1996).
\bibitem{TS}
M.~Takahashi and M.~Suzuki,
{\it Prog.\ Theor.\ Phys.} {\bf 48}, 2187 (1972).
\bibitem{mel}
E.~Melzer, 
{\it Int.\ J. Mod.\ Phys.\ A} {\bf 9}, 1115 (1994).
\bibitem{Chim} 
L.~Chim, ``Central Charge and the Andrews-Bailey Construction'', 
hep-th/9607168.
\bibitem{KP84}
V.~G.~Kac and D.~H.~Peterson,
{\it Adv.\ in Math.} {\bf 53}, 125 (1984).
\bibitem{JM84}
M.~Jimbo and T.~Miwa,
{\it Adv.\ Stud.\ in Pure Math.} {\bf 4}, 97 (1984).
\bibitem{LP85}
J.~Lepowsky and M.~Primc,
``Structure of the standard modules of the affine Lie algebra A$_1^{(1)}$'',
{\it Contemporary Mathematics}, vol. 46 (AMS, Providence, 1985).
\end{thebibliography}
\end{document}